\documentclass[prl,preprint,superscriptaddress,showpacs,amsmath,amssymb,aps]{revtex4-1}

\usepackage{graphicx}
\usepackage{amsmath}
\usepackage[version=3]{mhchem}
\usepackage[usenames]{color}
\usepackage{color}
\usepackage{physics}
\usepackage[colorlinks=true, linkcolor=blue, citecolor=blue, urlcolor=blue,pdftex]{hyperref}
\definecolor{newcolor}{rgb}{0.9,0,0.1}

\usepackage{orcidlink}
\usepackage{xr}

\newcommand{\figref}[1]{Fig.~\ref{#1}}

\begin{document}

\title[MoS2 Exciton on hBN]{Atomically-resolved exciton emission from single defects in MoS\textsubscript{2}}

\author{Lysander Huberich}
\affiliation{nanotech@surfaces Laboratory, Empa -- Swiss Federal Laboratories for Materials Science and Technology, D\"ubendorf 8600, Switzerland}

\author{Eve Ammerman}
\affiliation{nanotech@surfaces Laboratory, Empa -- Swiss Federal Laboratories for Materials Science and Technology, D\"ubendorf 8600, Switzerland}

\author{Gu Yu}
\affiliation{Key Laboratory of Artificial Structures and Quantum Control (Ministry of Education), Tsung-Dao Lee Institute, School of Physics and Astronomy, Shanghai Jiao Tong University, Shanghai, 200240, China}

\author{Yining Ren}
\affiliation{Key Laboratory of Artificial Structures and Quantum Control (Ministry of Education), Tsung-Dao Lee Institute, School of Physics and Astronomy, Shanghai Jiao Tong University, Shanghai, 200240, China}

\author{Sotirios Papadopoulos\,\orcidlink{0000-0002-3225-8239}}
\affiliation{Photonics Laboratory, ETH Zürich, Zürich 8093, Switzerland}
\affiliation{Universit\'e de  Strasbourg, CNRS, Institut de Physique et Chimie des  Mat\'erieux de Strasbourg, UMR 7504, F-67000 Strasbourg, France}

\author{Chengye Dong}
\affiliation{Two-Dimensional Crystal Consortium, The Pennsylvania State University, University Park, PA 16802, USA}

\author{Joshua A. Robinson\,\orcidlink{0000-0002-1513-7187}}
\affiliation{Department of Materials Science and Engineering, The Pennsylvania State University, University Park, PA 16082, USA}
\affiliation{Two-Dimensional Crystal Consortium, The Pennsylvania State University, University Park, PA 16802, USA}
\affiliation{Department of Chemistry and Department of Physics, The Pennsylvania State University, University Park, PA, 16802, USA}

\author{Kenji Watanabe\,\orcidlink{0000-0003-3701-8119}}
\affiliation{Research Center for Electronic and Optical Materials, National Institute for Materials Science, 1-1 Namiki, Tsukuba 305-0044, Japan}

\author{Takashi Taniguchi\,\orcidlink{0000-0002-1467-3105}}
\affiliation{Research Center for Materials Nanoarchitectonics, National Institute for Materials Science,  1-1 Namiki, Tsukuba 305-0044, Japan}

\author{Oliver Gröning}
\affiliation{nanotech@surfaces Laboratory, Empa -- Swiss Federal Laboratories for Materials Science and Technology, D\"ubendorf 8600, Switzerland}

\author{Lukas Novotny}
\affiliation{Photonics Laboratory, ETH Zürich, Zürich 8093, Switzerland}

\author{Tingxin Li\,\orcidlink{0000-0002-3572-4530}}
\affiliation{Key Laboratory of Artificial Structures and Quantum Control (Ministry of Education), Tsung-Dao Lee Institute, School of Physics and Astronomy, Shanghai Jiao Tong University, Shanghai, 200240, China}

\author{Shiyong Wang\,\orcidlink{0000-0001-6603-9926}}
\affiliation{Key Laboratory of Artificial Structures and Quantum Control (Ministry of Education), Tsung-Dao Lee Institute, School of Physics and Astronomy, Shanghai Jiao Tong University, Shanghai, 200240, China}

\author{Bruno Schuler\,\orcidlink{0000-0002-9641-0340}}
\email[]{bruno.schuler@empa.ch}
\affiliation{nanotech@surfaces Laboratory, Empa -- Swiss Federal Laboratories for Materials Science and Technology, D\"ubendorf 8600, Switzerland}

\begin{abstract} 
\textbf{
Understanding how atomic defects shape the nanoscale optical properties of two-dimensional (2D) semiconductors is essential for advancing quantum technologies and optoelectronics.  
Using scanning tunneling spectroscopy (STS) and luminescence (STML), we correlate the atomic structure and optical fingerprints of individual defects in monolayer MoS$_2$. 
A bilayer of hexagonal boron nitride (hBN) effectively decouples MoS$_2$ from the graphene substrate, increasing its band gap and extending the defect charge state lifetime. This enables the observation of sharp STML emission lines from MoS$_2$ excitons and trions exhibiting nanoscale sensitivity to local potential fluctuations. We identify the optical signatures of common point defects in MoS$_2$: sulfur vacancies (Vac$_\text{S}^-$), oxygen substitutions (O$_\text{S}$), and negatively charged carbon-hydrogen complexes (CH$_\text{S}^-$). While Vac$_\text{S}^-$ and O$_\text{S}$ only suppress pristine excitonic emission, CH$_\text{S}^-$ generate defect-bound exciton complexes ($A^-X$) about 200\,meV below the MoS$_2$ exciton. Sub-nanometer-resolved STML maps reveal large spectral shifts near charged defects, concurrent with the local band bending expected for band-to-defect optical transitions. These results establish an atomically precise correlation between structure, electronic states, and optical response, enabling deterministic engineering of quantum emitters in 2D materials.
}
\end{abstract}


\date{\today}
\maketitle

\section*{Introduction}\label{sec:intro}
The unique optical and electronic properties of atomic defects in solid-state materials can be exploited as an enabling technology that underpin applications in quantum sensing, communication, and integrated photonics~\cite{aharonovichSolidstatesinglephotonemitters2016b,atatureMaterialPlatformsSpinbased2018a,wolfowiczQuantumguidelinessolidstate2021}. Significant research efforts have been focused on exploring their exceptional properties, including their spin-photon interface~\cite{gruberScanningconfocaloptical1997,castellettosiliconcarbideroomtemperature2014a} and long coherence times~\cite{maurerRoomtemperaturequantumbit2012a}, and, as a prerequisite for their deterministic engineering, identifying the atomic origin of these defects. In recent years, two-dimensional (2D) materials have emerged as a promising platform for hosting quantum emitters. The absence of surface dangling bonds in 2D materials minimizes decoherence effects and enhances photon indistinguishability, addressing limitations faced by bulk materials~\cite{liu2Dmaterialsquantum2019,wolfowiczQuantumguidelinessolidstate2021}.
Furthermore, 2D semiconductors offer facile bottom-up incorporation of impurities, high light extraction efficiency, enable electric field tuning and spatial engineering of defects, and are compatible with on-chip integration~\cite{liu2Dmaterialsquantum2019,moody2022Roadmapintegrated2022,brotons-gisbertQuantumPhotonics2D2023}.\\

Defects in transition metal dichalcogenides (TMDs)~\cite{komsaTwoDimensionalTransitionMetal2012a,wangAtomicstructuredefects2018a,barjaIdentifyingSubstitutionalOxygen2019a,schulerHowSubstitutionalPoint2019a,schulerLargeSpinOrbitSplitting2019b} and hexagonal boron nitride (hBN)~\cite{tranQuantumemissionhexagonal2016,gottschollInitializationreadoutintrinsic2020,hayeeRevealingmultipleclasses2020a,mendelsonIdentifyingcarbonsource2020,Qiu:2024qyq} have been widely studied, and single-photon emission from these materials is frequently observed~\cite{Srivastava:2015zvh,He:2015pup,tranQuantumemissionhexagonal2016,grossoTunablehighpurityroom2017a,gottschollInitializationreadoutintrinsic2020,Parto:2020fmx}.
Different sources of single-photon emission have been suggested, such as mesoscopic strain fields~\cite{Srivastava:2015zvh,Parto:2020fmx}, and atomic defects including vacancies~\cite{kleinSiteselectivelygeneratedphoton2019a}, impurity atoms~\cite{mendelsonIdentifyingcarbonsource2020}, and defect complexes~\cite{sunUnveilingsulfurvacancy2024}.
Far-field optical studies have revealed substantial spatial variations in emitter properties, highlighting the critical role of the local environment, including strain, dielectric effects, potential fluctuation, and proximity to other defects.
However, directly correlating structural defects to their optical response remains a challenge since atomic-scale structural characterization and micrometer-scale far-field optical technique operate at vastly different length scales. 
Near-field techniques~\cite{baoVisualizingNanoscaleExcitonic2015,zhangNanospectroscopyExcitonsAtomically2022} and cathodoluminescence~\cite{hayeeRevealingmultipleclasses2020a,sunUnveilingsulfurvacancy2024} have reduced this gap but resolution remains limited to tens of nanometers.
Moreover, recent advances in controlled defect creation have started to address the lack of spatially resolved detection via impurity doping~\cite{mendelsonIdentifyingcarbonsource2020,lohNbimpurityboundexcitons2024} or ion irradiation~\cite{kleinSiteselectivelygeneratedphoton2019a,mitterreiterrolechalcogenvacancies2021}. Nonetheless, this inability to concurrently resolve atomic-scale structure and optical properties has remained a significant barrier to the unambiguous identification of quantum emitters and the underlying causes of their spectroscopic variability. 
\\

Here, we demonstrate atomic-scale correlation of structural, electronic, and optical properties in TMDs by means of scanning tunneling microscopy-induced luminescence (STML). We electronically decouple a MoS$_2$ monolayer from the substrate by a bilayer hBN. This decoupling is evident by an increase of the electronic TMD band gap and extended charge state lifetime of defects. The TMD/hBN/Gr heterostructure facilitates electrically-stimulated photon emission from pristine MoS$_2$ excitons and trions. Variations of the exciton-to-trion emission ratios, observable in spatially resolved spectral mapping, reveal nanoscale sensitivity to local potential fluctuations.
Importantly, we identify the spectral fingerprints of common point defects in MoS$_2$, including sulfur vacancies (Vac$_\text{S}^-$), oxygen substitutions (O$_\text{S}$), and negatively charged carbon-hydrogen complexes (CH$_\text{S}^-$).
While Vac$_\text{S}^-$ and O$_\text{S}$ suppress pristine exciton emission,  CH$_\text{S}^-$ hosts defect-bound excitons approximately 200\,meV below the MoS$_2$ exciton. The dependence of optical emission on injection bias and current reveals the charge-mediated excitation mechanism. The strong spectral shift of defect-bound excitons with tip position, following the local band bending around the charged impurity, is consistent with a band-to-defect transition from the conduction band to the localized acceptor state of CH$_\text{S}^-$.

\section*{Results and Discussion}

\subsection*{TMD on ultrathin hBN decoupling layers}
TMDs on graphene substrates enable local electronic characterization by scanning tunneling microscopy/spectroscopy (STM/STS), including the detection of in-gap defect states~\cite{schulerLargeSpinOrbitSplitting2019b,schulerHowSubstitutionalPoint2019a}, while preserving their direct band gap~\cite{ugedaGiantbandgaprenormalization2014}. However, TMD exciton emission is strongly quenched, owing to wavefunction hybridization and rapid non-radiative hot exciton transfer to graphene~\cite{lorchatFilteringphotoluminescencespectra2020a}.
To mitigate these effects, we introduce bilayer hBN between the TMD and graphene to suppress non-radiative decay channels while maintaining a low tunneling barrier for efficient charge injection~\cite{nisiScanningprobespectroscopy2024}.
Moreover, the decoupling layer increases the charge carrier lifetime and allows for electrically stimulated exciton formation by STML~\cite{doppagneElectrofluorochromismsinglemoleculelevel2018}.
We choose MoS$_2$ because it exhibits favorable asymmetric level alignment of the band edges with respect to the Fermi level of graphene. This ensures that holes injected into the valence band provide sufficient energy to drive charge-mediated exciton formation~\cite{miwaManyBodyStateDescription2019a,jiangManyBodyDescriptionSTMInduced2023a}.\\

\begin{figure*}[h]
\includegraphics[width=\textwidth]{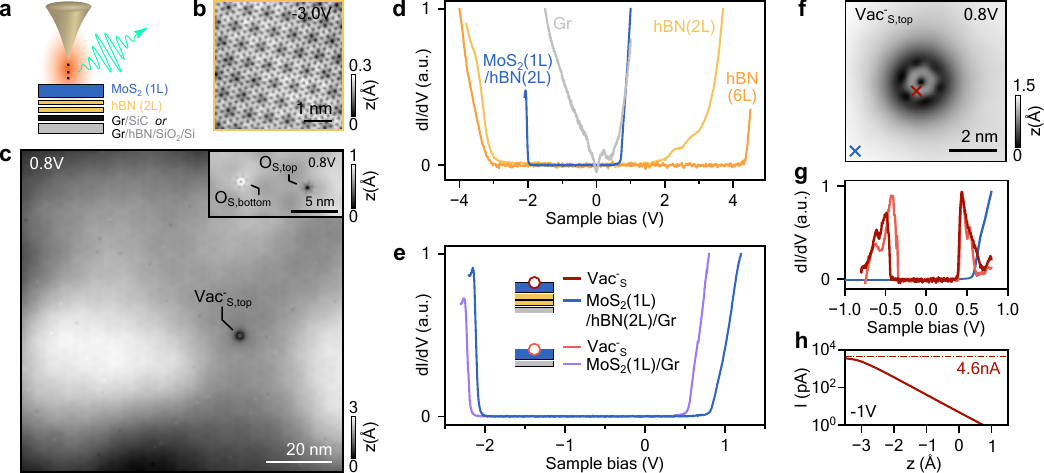}
\caption{\label{fig:fig1}
\textbf{Electronic property of MoS$_2$ on ultrathin hBN decoupling layers.}
\textbf{a}, Sketch of electrically-stimulated MoS$_2$(1L)/hBN(2L)/Gr heterostructure. Both quasi-freestanding epitaxial graphene on SiC and few-layer graphene on hBN/SiO$_2$/Si have been used as substrates, denoted as Gr for simplicity. 
\textbf{b}, STM topography of bilayer hBN on Gr moiré pattern.
\textbf{c}, Large scale STM topography of MoS$_2$(1L)/hBN(2L)/Gr heterostructure with low density of intrinsic isovalent oxygen substitutions (O$_\text{S,top/bottom}$, see inset) and an ion beam-induced negatively charged sulfur vacancy (Vac$_\text{S,top}^-$).
\textbf{d}, STS of multilayer (6L) hBN, bilayer hBN, monolayer MoS$_2$/hBN(2L), and epitaxial graphene on SiC.
\textbf{e}, Comparison of band gap and band alignment of MoS$_2$/Gr and MoS$_2$/hBN(2L)/Gr.
\textbf{f}, STM image of Vac$_\text{S,top}^-$ on MoS$_2$/hBN(2L)/Gr with STS locations from (g) indicated.
\textbf{g}, Comparison of Vac$_\text{S,top}^-$ defect states on  MoS$_2$/Gr (light red) and  MoS$_2$/hBN(2L)/Gr (dark red), respectively.
\textbf{h}, Z (height) spectroscopy ($V = -1\,$V) on Vac$_\text{S,top}^-$/MoS$_2$/hBN(2L)/Gr exhibiting current saturation behavior.
}
\end{figure*}

\figref{fig:fig1}a illustrates the heterostructure, consisting of a monolayer MoS$_2$ on bilayer hBN, supported by either quasi-freestanding epitaxial graphene on SiC or few-layer graphene on hBN/SiO$_2$/Si substrates, hereafter referred to as Gr for simplicity.
The heterostructure was fabricated by the PC/PDMS dry-pickup and transfer method. Flake thickness and quality were validated via optical and atomic force microscopy prior to assembly. Post-transfer AFM cleaning of the heterostructure yields an ultraclean sample surface illustrated in the large-scale STM topography in \figref{fig:fig1}c. STS spectra are used to distinguish different surface materials, including graphene, bilayer hBN, multilayer hBN, and monolayer MoS$_2$, as shown in \figref{fig:fig1}d.
The absence of a band gap for Gr and band edges for MoS$_2$ is clearly observed. The hBN band onsets in STS measurements exhibit strong layer dependence (see Fig. S4). Notably, as the hBN thickness increases, the conduction band edge shifts to higher energies, whereas the valence band edge remains nearly unchanged.
Beyond STS, STM imaging is crucial for identifying surface materials. hBN often exhibits moiré superlattices due to lattice mismatch and twist angles with Gr (\figref{fig:fig1}b). 
MoS$_2$ features characteristic point defects previously identified~\cite{barjaIdentifyingSubstitutionalOxygen2019a,schulerHowSubstitutionalPoint2019a}, distinguishing it from hBN and graphene. The defect density in as-exfoliated MoS$_2$ is very low ($6.4 \times 10^{-11}\,\text{cm}^{-2}$), consisting primarily of oxygen substitutions (O$_\text{S}$) (\figref{fig:fig1}c inset). Here we deliberately introduce sulfur vacancies (Vac$_\text{S}^-$)~\cite{schulerLargeSpinOrbitSplitting2019b}, and carbon-hydrogen complexes (CH$_\text{S}^-$)~\cite{cochraneIntentionalcarbondoping2020b}, through mild argon sputtering (see methods). 
The latter are common defects in synthetically grown TMDs~\cite{cochraneIntentionalcarbondoping2020b}.
\\

Introducing bilayer hBN between MoS$_2$ and Gr, leads to a pronounced upward shift of the MoS$_2$ band onsets due to the higher substrate work function and increased band gap by approximately 30\,meV due to reduced screening (\figref{fig:fig1}e and S5d)~\cite{ugedaGiantbandgaprenormalization2014,riis-jensenElectricallycontrolleddielectric2020}. 
The Vac$_\text{S}^-$ STM topography (\figref{fig:fig1}f) and defect states (\figref{fig:fig1}g) retain their character, but the HOMO--LUMO splitting is slightly increased on hBN supported TMD compared to their non-decoupled counterparts. 
On both substrates, the sulfur vacancy is negatively charged, as evident by the dark halo in the STM topography, due to the specific Fermi level alignment with the underlying graphene substrate~ \cite{mitterreiterAtomisticPositioningDefects2020a,xiangChargestatedependentsymmetry2024a}.
Z (height) spectroscopy measurements show a current saturation (\figref{fig:fig1}h) by probing the sequential tunneling from the defect state to the substrate, further confirm the enhanced decoupling of hBN supported MoS$_2$. While Vac$_\text{S}^-$/MoS$_2$(1L)/Gr show saturation currents exceeding 100\,nA (implying carrier lifetimes on the order of 1\,ps)~\cite{bobzienLayerDependentChargeStateLifetime2025a}, Vac$_\text{S}^-$/MoS$_2$(1L)/\textit{hBN(2L)}/Gr exhibits a reduced saturation current of about 5\,nA, corresponding to an enhanced charge state lifetime of about 30\,ps. The increased band gap of both MoS$_2$ and Vac$_\text{S}^-$, along with the significantly extended charge state lifetime, indicate the effective electronic decoupling of MoS$_2$ from the graphene substrate facilitated by the bilayer hBN.
\\

\subsection*{Exciton emission of pristine MoS$_2$ on ultrathin hBN}

\begin{figure*}
\includegraphics[width=\textwidth]{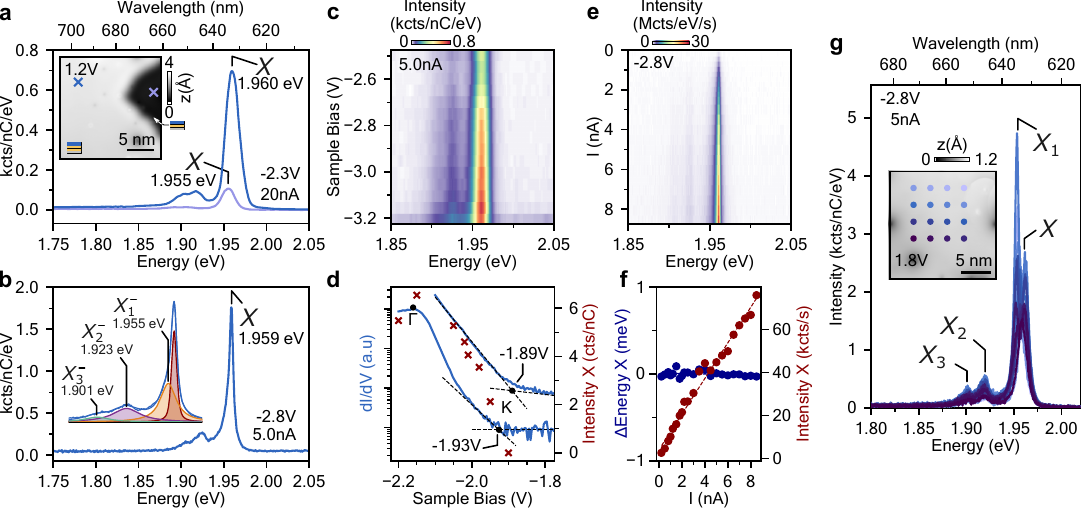}
\caption{\label{fig:fig2}
\textbf{Exciton and trion emission of pristine MoS$_2$.}
\textbf{a}, STML emission from monolayer MoS$_2$ supported by mono- (blue) and bilayer (purple) hBN, respectively (150\,l/mm grating). The spectra location are indicated in the STM topography inset.
\textbf{b}, STML emission from pristine MoS$_2$(1L)/hBN(2L)/Gr with a 600\,l/mm grating. The exciton $X$ and three trion $X_{1-3}^-$ emission energies extracted from Pseudo-Voigt fits are indicated.
\textbf{c}, Voltage dependence of STML emission (150\,l/mm grating). 
\textbf{d}, Valence band edge from d$I$/d$V$ spectra overlaid with the exciton emission intensity (red crosses).
\textbf{e}, Current dependence of STML emission (600\,l/mm grating).
\textbf{f}, Exciton peak energy shift and emission intensity as a function of excitation current. The dashed lines represent linear fits.
\textbf{g}, STML emission spectra measured on a 4$\times$4 grid near defects (spectra locations in inset) exhibit variations in the exciton-to-trion emission ratio (600\,l/mm grating).
The data shown in c-g are recorded on MoS$_2$(1L)/hBN(2L)/Gr.}
\end{figure*}

Electroluminescence in vertical tunnel devices~\cite{withersLightemittingdiodesbandstructure2015a,palacios-berraqueroAtomicallythinquantum2016,wangExcitonassistedelectrontunnelling2023} and STM geometries~\cite{pommierScanningTunnelingMicroscopeInduced2019,schulerElectricallydrivenphoton2020a,romanTunnelingcurrentinducedlocalexcitonic2020,lopezTipinducedexcitonicluminescence2023,gengImagingValleyExcitons2024} has been demonstrated previously. However, atomic-scale investigation of excitonic emission was so far precluded by non-cryogenic operation temperatures, insufficient sample quality, or lack of substrate decoupling.
Here we show that both mono- and bilayer hBN provides sufficient decoupling of monolayer MoS$_2$ from Gr to enable electrically stimulated exciton emission in STML experiments (see Fig.~\ref{fig:fig2}a). Notably, the brightness of STML emission is enhanced sevenfold, when MoS$_2$ is supported by bilayer hBN rather than monolayer hBN.
This enhancement is accompanied by a slight inrease in emission energy for MoS$_2$(1L)/hBN(2L) attributable to the reduced screening from the graphene substrate, consistent with the increased electronic band gap observed in STS (see Fig. S5).\\

Fig.~\ref{fig:fig2}b presents a high-resolution STML spectrum of monolayer MoS$_2$ supported by bilayer hBN on Gr. We observe a sharp emission line centered at 1.959\,eV, accompanied by three smaller side-peaks at 1.955\,eV, 1.923\,eV, and 1.901\,eV, respectively. For samples with no or minimal ion irradiation, STML spectra exhibit minimal spatial variations.
The STML measurement closely resembles low-temperature photoluminescence (PL) spectra of high-quality MoS$_2$ samples~\cite{qianProbingDarkExcitons2024}.
Based on direct correspondence with these literature data, we tentatively assign the main emission peaks to the pristine MoS$_2$ exciton $X$ and three trion species, $X_{1,2,3}^-$, as labeled in \figref{fig:fig2}b.
The assignment of the low energy peaks in the PL fine-structure of MoS$_2$ is complicated by the small conduction band splitting and susceptibility to many-body effects~\cite{druppelDiversitytrionstates2017}. 
However, recent magneto-PL measurements and time-dependent density functional theory (TD-DFT) calculations converge on three trion configurations involving mixed intravalley and intervalley singlet and triplet states~\cite{qianProbingDarkExcitons2024,druppelDiversitytrionstates2017,grzeszczykExposingtrionsfine2021,kleinTrions$mathrmMoS_2$are2022}.\\

The STML emission is mediated by tunneling into the valence band, where the emission onset at $-1.9\,$V coincides with valence band maximum (\figref{fig:fig2}d).
No excitonic emission was observed at positive sample bias. However, at high positive bias around 3.5\,V, inelastic transitions into unoccupied MoS$_2$ states drive broadband plasmon emission, which exhibits a narrow absorption dip at the exciton energy (see Fig. S10).
The emission intensity scales linearly with the tunneling current as expected for a single electron process, and exhibits minimal spectral shifts of the exciton line (\figref{fig:fig2}e,f).
However, the peak intensities fluctuate on a nanometer length scale in the vicinity of charged defects. Fig.~\ref{fig:fig2}g shows STML spectra of MoS$_2$(1L)/hBN(2L) at various positions, neighboring charged defects visible as dark depressions in the STM topography. 
This variability is particularly pronounced for the $X_{1}^-$ trion. 
While such charged defects are artificially introduced in our exfoliated samples, they are commonly found in synthetic TMDs~\cite{schulerHowSubstitutionalPoint2019a,cochraneIntentionalcarbondoping2020b}.

\subsection*{Atomically-resolved defect exciton emission}
In the previous section we present how proximity to defects influences the emission properties of pristine MoS$_2$. Next, we focus on the local emission characteristics of specific point defects, namely sulfur vacancies (Vac$_\text{S}^-$), oxygen substitutions (O$_\text{S}$), and negatively charged CH$_\text{S}^-$. 
In as-exfoliated samples, only O$_\text{S}$ and other isovalent substitutional defects are initially present (\figref{fig:fig1}c inset). Vac$_\text{S}^-$ and CH$_\text{S}^-$ defects are deliberately introduced via mild argon sputtering under ultrahigh vacuum.
While argon sputtering of MOCVD-grown samples leads mostly to S vacancy creation, on the exfoliated samples both Vac$_\text{S}^-$ and CH$_\text{S}^-$ are observed~\cite{mitterreiterAtomisticPositioningDefects2020a}. 
We attribute the formation of CH$_\text{S}^-$ to the decoration of reactive vacancy sites by residual polymers or other organic species from flake transfer. 
CH$_\text{S}^-$ are ionized acceptor-type defects that carry a negative charge and induce strong upward band bending~\cite{schulerHowSubstitutionalPoint2019a,aghajanianResonantboundstates2020,cochraneIntentionalcarbondoping2020b}. Although absent in as-exfoliated samples, CH$_\text{S}^-$ are common in as-grown CVD and MOCVD TMD samples~\cite{schulerHowSubstitutionalPoint2019a,cochraneIntentionalcarbondoping2020b}.\\

\begin{figure*}[h]
\includegraphics[width=0.5\textwidth]{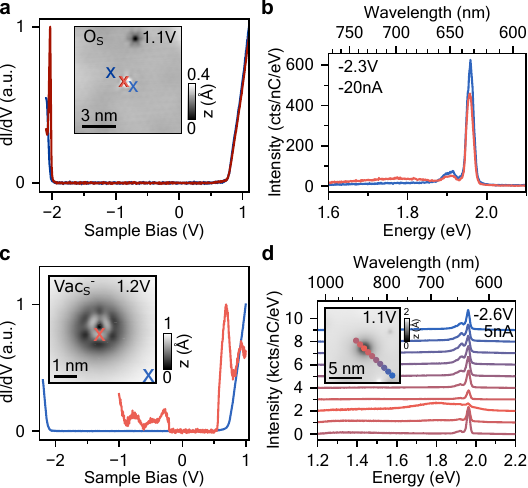}
\caption{\label{fig:fig3}
\textbf{Suppressed STML emission at Vac$_{\text{S}}^{-}$ and O$_{\text{S}}$.}
\textbf{a}, STS spectra on top of the O$_\text{S}$ defect (orange) and pristine MoS$_2$(1L)/hBN(2L)/Gr (blue) revealing a resonant defect state at -2\,V, but no in-gap defect state. The STM topography inset illustrates the locations of STS measurements and STML spectra shown in panel (b).
\textbf{b}, STML emission recorded directly above and adjacent to the O$_\text{S}$ defect, highlighting suppression of pristine emission due to the defect. 
\textbf{c}, STS spectra of Vac$^-_{\text{S}}$ and pristine MoS$_2$(1L)/hBN(2L)/Gr, at the positions indicated in the inset. 
\textbf{d}, Spatially resolved STML emission along a line across Vac$^-_{\text{S}}$, showing a reduction in the pristine emission intensity as the tip approaches the defect. A broadband plasmonic emission is observed when the tip is positioned directly above Vac$^-_{\text{S}}$.
}
\end{figure*}

In \figref{fig:fig3}a,b STS and STML across an O$_\text{S}$ defect are shown. No additional emission lines are observed at the defect site due to the absence of in-gap defect states.
However, a reduction in the intensity of the pristine exciton and trion emission is evident, suggesting additional non-radiative decay channels. On graphene substrates, the sulfur vacancy in MoS$_2$ is negatively charged, as evident by the dark halo in the STM topography~ \cite{mitterreiterAtomisticPositioningDefects2020a,xiangChargestatedependentsymmetry2024a}. For Vac$_\text{S}^-$ defects, often proposed as origins of single photon emission in far-field optical measurements~\cite{mitterreiterrolechalcogenvacancies2021}, our results are surprising. 
Although Vac$_\text{S}^-$ in MoS$_2$(1L)/hBN(2L) exhibits distinct in-gap defect states (\figref{fig:fig3}c), as previously reported for MoS$_2$(1L)/Gr~\cite{xiangChargestatedependentsymmetry2024a}, no excitonic emission lines emerge at the defect site (\figref{fig:fig3}d) within our detectable energy range ($>1.1$\,eV). Rather only broad plasmonic emission can be observed and no defect-specific absorption dips are visible (Fig. S10d). Like O$_\text{S}$, Vac$_\text{S}^-$ primarily suppresses the pristine emission.\\

\begin{figure*}[h]
\includegraphics[width=\textwidth]{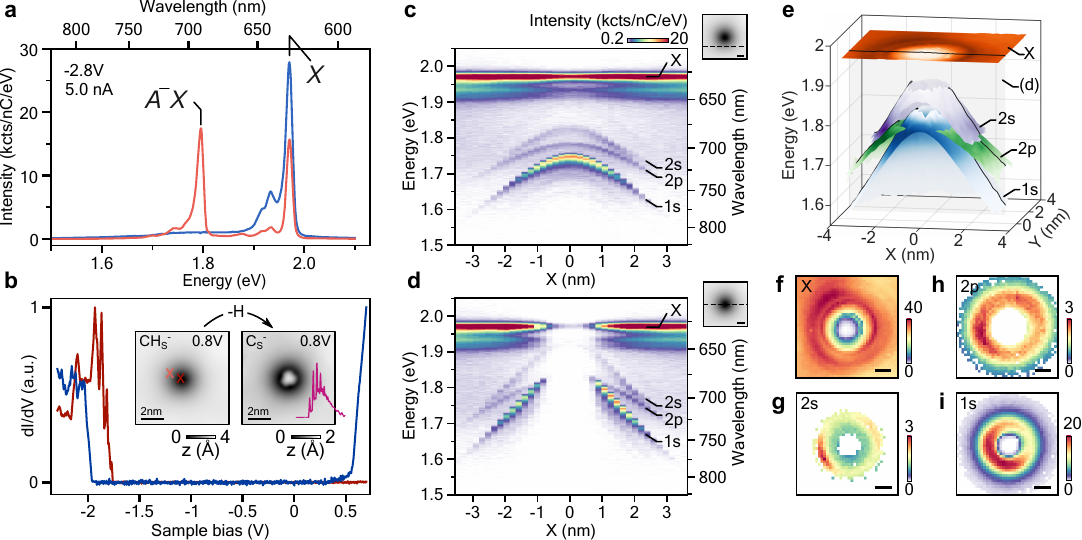}
\caption{\label{fig:fig4}
\textbf{Defect-bound exciton emission at negatively charged CH$^-_\text{S}$ defect.}
\textbf{a}, STML spectra of pristine MoS$_2$ (blue) and negatively charged CH$^-_\text{S}$ defect (orange) recorded at -2.8\,V and 5\,nA.  
\textbf{b}, STS spectra of CH$^-_\text{S}$ (red) and pristine MoS$_{2}$ (blue). The left inset shows the locations of the STML and STS measurements. The identity of the CH$^-_\text{S}$ defect is verified by cleaving off the H with a voltage pulse (4.5\,V, 8\,nA), creating C$^-_\text{S}$ with its characteristic vibronic spectrum (right inset)~\cite{cochraneSpindependentvibronicresponse2021c}.
\textbf{c,d}, STML line spectra off-center (c) and on-center (d) across CH$^-_\text{S}$ (-2.8\,V and 7.5\,nA), at the location indicated on the top right. While the MoS$_2$ exciton and trion energies stay constant, the CH$^-_\text{S}$ defect-bound excitons, labelled 1s, 2p, 2s, blue-shift significantly near the defect center.
\textbf{e}, 3D plot of X (orange), 2s (purple), 2p (green) and 1s (blue) exciton peaks as a function of energy and spatial coordinate with relative peak intensity indicated by the colorscale. The shaded plane and black lines mark the cross-section shown in (d).
\textbf{f-i}, STML intensity maps of the MoS$_2$ exciton ($X$, f) and CH$^-_\text{S}$ excitons (g-i). The map shows the emission peak intensity at each point corrected for shifts in the emission line. Scale bar: 1\,nm. Colorscale: kcts/nC/eV.
}
\end{figure*}

In contrast, CH$_\text{S}^-$ exhibits distinct defect emission lines, as shown in \figref{fig:fig4}a. The main peak appears around 1.8\,eV, approximately 200\,meV below the MoS$_2$ exciton, concurrent with the suppression of the pristine STML emission. Interestingly, the spectral STML response of CH$_\text{S}^-$ resembles the far-field PL signature often attributed to sulfur vacancies~\cite{mitterreiterrolechalcogenvacancies2021}. 
The d$I$/d$V$ spectrum of the same CH$_\text{S}^-$ defect (Fig.~\ref{fig:fig4}b), reveals characteristic acceptor states and band bending, consistent with previous reports~\cite{aghajanianResonantboundstates2020}. The chemical identity of the CH$_\text{S}^-$ defect was confirmed by cleaving off the H with a voltage pulse, creating C$^-_\text{S}$ with its characteristic vibronic spectrum~\cite{cochraneSpindependentvibronicresponse2021c}.\\

Fig.~\ref{fig:fig4}c,d shows sub-nanometer resolved STML spectra acquired across and slightly off the CH$_\text{S}^-$ defect center. While the emission energies of the pristine exciton and trions remain virtually unchanged, the defect emission lines exhibit a pronounced blue shift as the STM tip moves closer to the defect center.
We attribute the strong spectral blue shifts to the local band bending induced by the negative charge localized at the defect, as further argued in the next section. 
In addition to the main defect emission line (labeled 1s in Fig.~\ref{fig:fig4}c,d), two higher-energy features can be identified and labeled 2p and 2s in analogy to the hydrogenic Rydberg series~\cite{chernikovExcitonBindingEnergy2014a}. 
Interestingly, while the 1s and 2s lines have a similar curvature, 2p follows a different trajectory. Faint sidebands on the low-energy flank of the 1s line are attributed to phonon replicas. In Fig.~\ref{fig:fig4}e, the 3D reconstruction of defect emission peaks extracted from the hyperspectral dataset is shown as a function of emission energy and spatial coordinate.\\ 

Fig.~\ref{fig:fig4}f-i display photon maps of the intrinsic MoS$_2$ exciton and the CH$_\text{S}^-$-bound defect excitons corrected for spectral shifts.
The MoS$_2$ exciton intensity is slightly enhanced at a distance of 2.2\,nm from the defect but fully suppressed at the defect center. By contrast, defect-bound excitons form localized ring-like emission patterns with maxima at 1.2\,nm (1s), 1.4\,nm (2s), and 1.7\,nm (2p). This is consistent with the expected exciton Bohr radius in TMDs, which is on the order of a few nanometers.~\cite{wangColloquiumExcitonsAtomically2018b}. The emission maps do not directly reflect the intrinsic exciton transition density, as the signal is convoluted with the vertical tip dipole and associated luminescence quenching. The central suppression of defect emission is consistent with a perpendicular alignment between the in-plane transition dipole of the emitter and the vertical tip dipole, which cancels the signal. S-wave excitons, with finite wavefunction amplitude at the defect center, couple more strongly to the localized charge than p-type excitons, which exhibit a nodal structure. This interpretation is consistent with the observed photon maps. \\

Finally, the non-hydrogenic Rydberg series of Wannier–Mott excitons in TMDs~\cite{chernikovExcitonBindingEnergy2014a} is known to exhibit lifted degeneracy between states of equal principal quantum number~\cite{yeProbingexcitonicdark2014}. The fact that the 2p state lies energetically below the 2s state is a hallmark of the non-local dielectric screening described by the Rytova–Keldysh potential in two dimensions. The following section addresses the microscopic character of the defect-bound exciton and the charge-mediated excitation mechanism.

\subsection*{Character of defect-bound exciton and electrical excitation pathway \label{sec:theory}}
\begin{figure*}[h]
\includegraphics[width=0.8\textwidth]{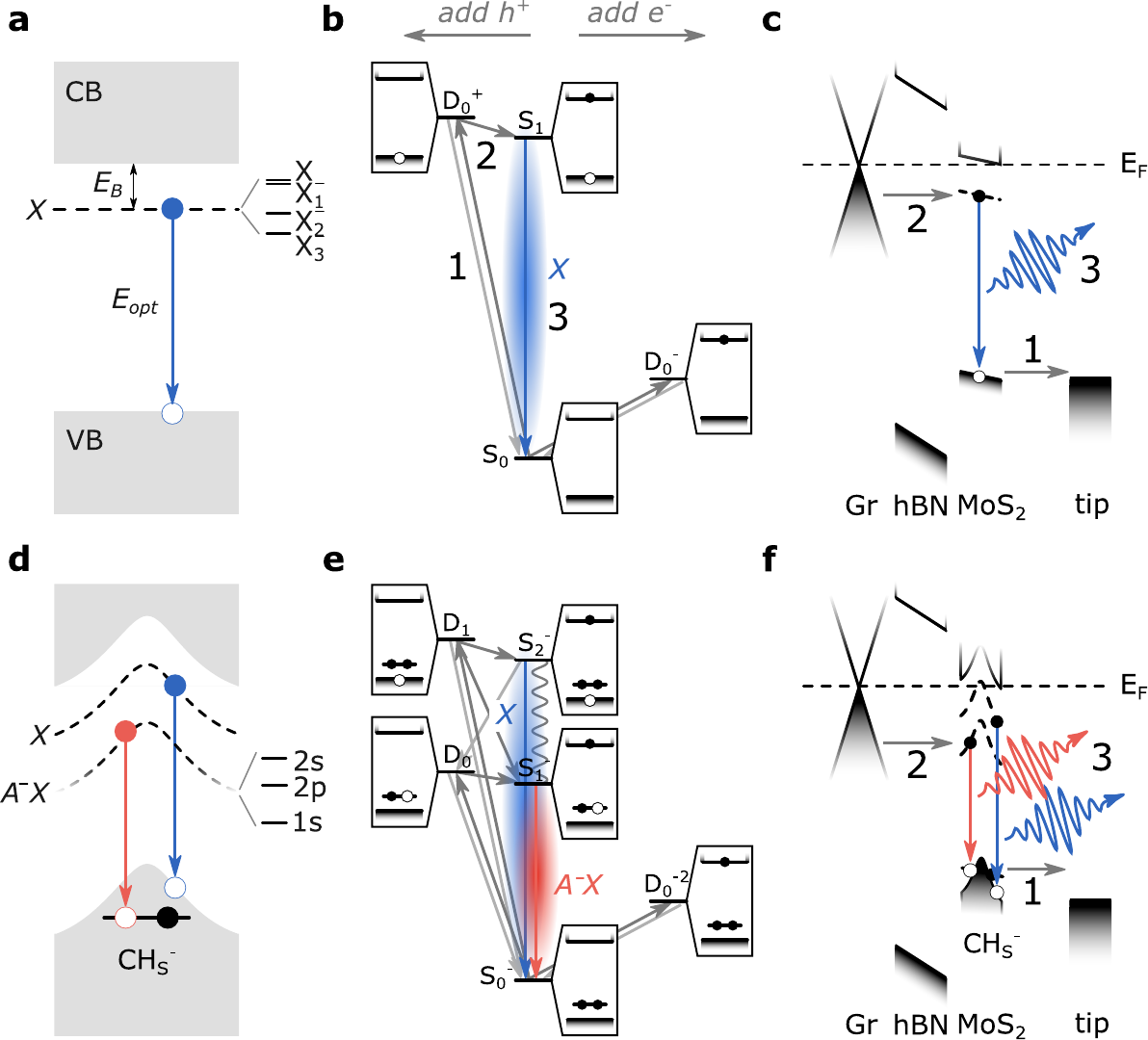}
\caption{\label{fig:fig5}
\textbf{Electroluminescence mechanism for pristine MoS$_2$ and defect-bound excitons localized at CH$_\text{S}^-$.}
\textbf{a}, Schematic level diagram of the MoS$_2$ quasi-particle bands and exciton $X$ and trion $X_{1-3}^-$ energies as derived from STS and STML spectra. $E_\text{B}$ is the exciton binding and $E_\text{opt}$ the emission energy, respectively. \textbf{b}, Many-body level diagram of the pristine MoS$_2$ STML emission via hole attachment to the valence band. \textbf{c}, Corresponding STM double barrier tunneling junction indicating hole attachment (1), neutralization via electron tunneling from the substrate into the excitonic state (2), and radiative recombination (3). \textbf{d}, Schematic level diagram at a negatively charged CH$_\text{S}^-$ defect, with the band bending, band-to-band (blue) and band-to-defect transitions (orange) indicated. \textbf{e}, Many-body level diagram of the $X$ and defect-bound exciton emission $A^{-}X$ via transient hole attachment. \textbf{f}, Corresponding tunneling junction model.
}
\end{figure*}

STML emission from MoS$_2$ has been previously reported ~\cite{lopezTipinducedexcitonicluminescence2023} and follows the established excitation pathways of STM-induced light emission~\cite{miwaManyBodyStateDescription2019a,jiangManyBodyDescriptionSTMInduced2023a}, as schematically illustrated in \figref{fig:fig5}a-c. Hole attachment at negative sample bias promotes the sample from its ground state S$_0$ to the cationic state D$_0^+$ (S$_0$ $\rightarrow$ D$_0^+$; step 1). This excited state can neutralize either by refilling the valence band hole or by injecting an electron into the conduction band, thus reaching the neutral excited state S$_1$ (D$_0^+$ $\rightarrow$ S$_1$; step 2).
Owing to the exciton binding energy, electron injection into the bound exciton state occurs at energies significantly below the conduction band minimum. From the conduction and valence band onsets at 0.5\,V and -1.9\,V (see Fig. S6), we extract a transport gap of 2.4\,eV. Given the observed exciton emission at 1.95\,eV, this yields an exciton binding energy $E_\text{B}$ of approximately 450\,meV. From S$_1$, the exciton quickly recombines radiatively in the plasmonic nanocavity (S$_1$ $\rightarrow$ S$_0$; step 3). For simplicity, trion emission is not shown here.\\

In the presence of the negatively charged CH$_\text{S}^-$ defect, the energy landscape is significantly modified by a strong upward bending of both valence and conduction bands, as previously reported~\cite{schulerHowSubstitutionalPoint2019a,aghajanianResonantboundstates2020}.
In addition, the defect hosts an acceptor-like occupied defect state overlapping the valence band~\cite{cochraneIntentionalcarbondoping2020b}. In \figref{fig:fig5}d-f the excitation scheme for the defect emission is depicted. The band-to-band transition energy remains unchanged, as the conduction and valence bands experience equivalent shifts and exciton diffusion enables recombination to occur away from the charged defect. However, an additional optical channel arises from transitions between the conduction band and the localized defect state, denoted $A^-X$. This process is expected to follow the defect-induced band bending. Indeed, the spatial and spectral variations of the $A^-X$ line match the band bending measured by STS (Fig. S13). 
We have considered alternative explanations for the lateral spectral shifts, including a Lamb shift due to coupling with the strong plasmonic cavity, Stark shifts from the tip-sample electrostatic field, and metallic screening effects reducing the defect binding energy. However, all these effects would exhibit pronounced tip shape and/or tip height dependence (cf. Fig. S14), and typically result in energy shifts of only a few meV -- far smaller than the blue shifts of several hundred meV observed here.\\

We therefore conclude that the defect emission localized at the negatively charged CH$_\text{S}^-$ is best described as an exciton bound to an ionized acceptor-like impurity ($A^-X$), arising from a band-to-defect transition.
The optically excited defect state S$_1^-$ can be reached via tunneling into the valence band, direct injection into the defect state, or energy transfer from the pristine exciton (S$_2^- \rightarrow$ S$_1^-$). 
Such free-to-bound luminescence is well established for doped II-VI and III-V semiconductors~\cite{pankoveOpticalProcessesSemiconductors2012}, and theory predicts that negatively charged acceptors in TMDs can bind free excitons with large binding energies, forming stable bound-exciton complexes~\cite{mostaaniDiffusionquantumMonte2017a}.\\

\section*{Conclusions}
We have demonstrated atomically-resolved exciton emission from monolayer MoS$_2$ and individual atomic defects, decoupled by ultrathin hBN layers, using STML.
Bilayer hBN effectively isolates MoS$_2$ from the graphene substrate, increasing the TMD band gap and extending charge state lifetime of defect states to tens of picoseconds. 
This decoupling enables charge-mediated excitation of the pristine MoS$_2$ exciton ($X$) and multiple trion states ($X_{1,2,3}^-$), whose relative intensities exhibit nanoscale variations in proximity to local potential fluctuations induced by nearby defects.
We reveal that oxygen substitutions (O$_\text{S}$) and sulfur vacancies (Vac$_\text{S}^-$) suppress the prisitine MoS$_2$ emission, without exhibiting additional defect-related emission lines.  
In contrast, negatively charged CH$^-_\text{S}$ defects give rise to distinct defect-bound exciton emission ($A^-X$), approximately 200\,meV below the MoS$_2$ exciton. 
The optical signature of this acceptor-bound exciton complexes, previously observed in far-field photoluminescence, is now conclusively linked to its atomic origin.
Here we present spatial maps of the main and excited state defect-bound excitons with a spatial extent between one and two nanometer. 
The strong spectral blue shifts of the defect emission follows the energy profile of the local band bending at the negatively charged defect. We conclude that the defect emission localized at the negatively charged CH$_\text{S}^-$ is best characterized by an exciton bound to an ionized acceptor-like impurity ($A^-X$) resulting from a band-to-defect transition. 
By directly correlating local electronic structure with nanoscale optical response, this work provides fundamental insight into exciton–defect interactions in 2D semiconductors. The ability to resolve and assign single-defect optical fingerprints establishes a pathway toward the deterministic engineering of quantum emitters in 2D materials.

\section*{References}
\bibliographystyle{nature_bsc}
\bibliography{references.bib}

\begin{thebibliography}{6}%
\makeatletter
\providecommand \@ifxundefined [1]{%
 \@ifx{#1\undefined}
}%
\providecommand \@ifnum [1]{%
 \ifnum #1\expandafter \@firstoftwo
 \else \expandafter \@secondoftwo
 \fi
}%
\providecommand \@ifx [1]{%
 \ifx #1\expandafter \@firstoftwo
 \else \expandafter \@secondoftwo
 \fi
}%
\providecommand \natexlab [1]{#1}%
\providecommand \enquote  [1]{``#1''}%
\providecommand \bibnamefont  [1]{#1}%
\providecommand \bibfnamefont [1]{#1}%
\providecommand \citenamefont [1]{#1}%
\providecommand \href@noop [0]{\@secondoftwo}%
\providecommand \href [0]{\begingroup \@sanitize@url \@href}%
\providecommand \@href[1]{\@@startlink{#1}\@@href}%
\providecommand \@@href[1]{\endgroup#1\@@endlink}%
\providecommand \@sanitize@url [0]{\catcode `\\12\catcode `\$12\catcode `\&12\catcode `\#12\catcode `\^12\catcode `\_12\catcode `\%12\relax}%
\providecommand \@@startlink[1]{}%
\providecommand \@@endlink[0]{}%
\providecommand \url  [0]{\begingroup\@sanitize@url \@url }%
\providecommand \@url [1]{\endgroup\@href {#1}{\urlprefix }}%
\providecommand \urlprefix  [0]{URL }%
\providecommand \Eprint [0]{\href }%
\providecommand \doibase [0]{http://dx.doi.org/}%
\providecommand \selectlanguage [0]{\@gobble}%
\providecommand \bibinfo  [0]{\@secondoftwo}%
\providecommand \bibfield  [0]{\@secondoftwo}%
\providecommand \translation [1]{[#1]}%
\providecommand \BibitemOpen [0]{}%
\providecommand \bibitemStop [0]{}%
\providecommand \bibitemNoStop [0]{.\EOS\space}%
\providecommand \EOS [0]{\spacefactor3000\relax}%
\providecommand \BibitemShut  [1]{\csname bibitem#1\endcsname}%
\let\auto@bib@innerbib\@empty
\bibitem [{\citenamefont {Zomer}\ \emph {et~al.}(2014)\citenamefont {Zomer}, \citenamefont {Guimar{\~a}es}, \citenamefont {Brant}, \citenamefont {Tombros},\ and\ \citenamefont {{van Wees}}}]{zomerFastPickTechnique2014}%
  \BibitemOpen
  \bibfield  {author} {\bibinfo {author} {\bibfnamefont {P.~J.}\ \bibnamefont {Zomer}}, \bibinfo {author} {\bibfnamefont {M.~H.~D.}\ \bibnamefont {Guimar{\~a}es}}, \bibinfo {author} {\bibfnamefont {J.~C.}\ \bibnamefont {Brant}}, \bibinfo {author} {\bibfnamefont {N.}~\bibnamefont {Tombros}}, \ and\ \bibinfo {author} {\bibfnamefont {B.~J.}\ \bibnamefont {{van Wees}}},\ }\href {\doibase 10.1063/1.4886096} {\bibfield  {journal} {\bibinfo  {journal} {Applied Physics Letters}\ }\textbf {\bibinfo {volume} {105}},\ \bibinfo {pages} {013101} (\bibinfo {year} {2014})}\BibitemShut {NoStop}%
\bibitem [{\citenamefont {Ma}\ \emph {et~al.}(2013)\citenamefont {Ma}, \citenamefont {Odenthal}, \citenamefont {Mann}, \citenamefont {Le}, \citenamefont {Wang}, \citenamefont {Zhu}, \citenamefont {Chen}, \citenamefont {Sun}, \citenamefont {Yamaguchi}, \citenamefont {Tran}, \citenamefont {Wurch}, \citenamefont {McKinley}, \citenamefont {Wyrick}, \citenamefont {Magnone}, \citenamefont {Heinz}, \citenamefont {Rahman}, \citenamefont {Kawakami},\ and\ \citenamefont {Bartels}}]{maControlledArgonBeaminduced2013b}%
  \BibitemOpen
  \bibfield  {author} {\bibinfo {author} {\bibfnamefont {Q.}~\bibnamefont {Ma}}, \bibinfo {author} {\bibfnamefont {P.~M.}\ \bibnamefont {Odenthal}}, \bibinfo {author} {\bibfnamefont {J.}~\bibnamefont {Mann}}, \bibinfo {author} {\bibfnamefont {D.}~\bibnamefont {Le}}, \bibinfo {author} {\bibfnamefont {C.~S.}\ \bibnamefont {Wang}}, \bibinfo {author} {\bibfnamefont {Y.}~\bibnamefont {Zhu}}, \bibinfo {author} {\bibfnamefont {T.}~\bibnamefont {Chen}}, \bibinfo {author} {\bibfnamefont {D.}~\bibnamefont {Sun}}, \bibinfo {author} {\bibfnamefont {K.}~\bibnamefont {Yamaguchi}}, \bibinfo {author} {\bibfnamefont {T.}~\bibnamefont {Tran}}, \bibinfo {author} {\bibfnamefont {M.}~\bibnamefont {Wurch}}, \bibinfo {author} {\bibfnamefont {J.~L.}\ \bibnamefont {McKinley}}, \bibinfo {author} {\bibfnamefont {J.}~\bibnamefont {Wyrick}}, \bibinfo {author} {\bibfnamefont {K.}~\bibnamefont {Magnone}}, \bibinfo {author} {\bibfnamefont {T.~F.}\ \bibnamefont {Heinz}}, \bibinfo {author} {\bibfnamefont {T.~S.}\ \bibnamefont {Rahman}},
  \bibinfo {author} {\bibfnamefont {R.}~\bibnamefont {Kawakami}}, \ and\ \bibinfo {author} {\bibfnamefont {L.}~\bibnamefont {Bartels}},\ }\href {\doibase 10.1088/0953-8984/25/25/252201} {\bibfield  {journal} {\bibinfo  {journal} {Journal of Physics: Condensed Matter}\ }\textbf {\bibinfo {volume} {25}},\ \bibinfo {pages} {252201} (\bibinfo {year} {2013})}\BibitemShut {NoStop}%
\bibitem [{\citenamefont {Kazmierczak}\ \emph {et~al.}(2021)\citenamefont {Kazmierczak}, \citenamefont {Van~Winkle}, \citenamefont {Ophus}, \citenamefont {Bustillo}, \citenamefont {Carr}, \citenamefont {Brown}, \citenamefont {Ciston}, \citenamefont {Taniguchi}, \citenamefont {Watanabe},\ and\ \citenamefont {Bediako}}]{kazmierczakStrainFieldsTwisted2021}%
  \BibitemOpen
  \bibfield  {author} {\bibinfo {author} {\bibfnamefont {N.~P.}\ \bibnamefont {Kazmierczak}}, \bibinfo {author} {\bibfnamefont {M.}~\bibnamefont {Van~Winkle}}, \bibinfo {author} {\bibfnamefont {C.}~\bibnamefont {Ophus}}, \bibinfo {author} {\bibfnamefont {K.~C.}\ \bibnamefont {Bustillo}}, \bibinfo {author} {\bibfnamefont {S.}~\bibnamefont {Carr}}, \bibinfo {author} {\bibfnamefont {H.~G.}\ \bibnamefont {Brown}}, \bibinfo {author} {\bibfnamefont {J.}~\bibnamefont {Ciston}}, \bibinfo {author} {\bibfnamefont {T.}~\bibnamefont {Taniguchi}}, \bibinfo {author} {\bibfnamefont {K.}~\bibnamefont {Watanabe}}, \ and\ \bibinfo {author} {\bibfnamefont {D.~K.}\ \bibnamefont {Bediako}},\ }\href {\doibase 10.1038/s41563-021-00973-w} {\bibfield  {journal} {\bibinfo  {journal} {Nature Materials}\ }\textbf {\bibinfo {volume} {20}},\ \bibinfo {pages} {956} (\bibinfo {year} {2021})}\BibitemShut {NoStop}%
\bibitem [{\citenamefont {Yoo}\ \emph {et~al.}(2019)\citenamefont {Yoo}, \citenamefont {Engelke}, \citenamefont {Carr}, \citenamefont {Fang}, \citenamefont {Zhang}, \citenamefont {Cazeaux}, \citenamefont {Sung}, \citenamefont {Hovden}, \citenamefont {Tsen}, \citenamefont {Taniguchi}, \citenamefont {Watanabe}, \citenamefont {Yi}, \citenamefont {Kim}, \citenamefont {Luskin}, \citenamefont {Tadmor}, \citenamefont {Kaxiras},\ and\ \citenamefont {Kim}}]{yooAtomicElectronicReconstruction2019}%
  \BibitemOpen
  \bibfield  {author} {\bibinfo {author} {\bibfnamefont {H.}~\bibnamefont {Yoo}}, \bibinfo {author} {\bibfnamefont {R.}~\bibnamefont {Engelke}}, \bibinfo {author} {\bibfnamefont {S.}~\bibnamefont {Carr}}, \bibinfo {author} {\bibfnamefont {S.}~\bibnamefont {Fang}}, \bibinfo {author} {\bibfnamefont {K.}~\bibnamefont {Zhang}}, \bibinfo {author} {\bibfnamefont {P.}~\bibnamefont {Cazeaux}}, \bibinfo {author} {\bibfnamefont {S.~H.}\ \bibnamefont {Sung}}, \bibinfo {author} {\bibfnamefont {R.}~\bibnamefont {Hovden}}, \bibinfo {author} {\bibfnamefont {A.~W.}\ \bibnamefont {Tsen}}, \bibinfo {author} {\bibfnamefont {T.}~\bibnamefont {Taniguchi}}, \bibinfo {author} {\bibfnamefont {K.}~\bibnamefont {Watanabe}}, \bibinfo {author} {\bibfnamefont {G.-C.}\ \bibnamefont {Yi}}, \bibinfo {author} {\bibfnamefont {M.}~\bibnamefont {Kim}}, \bibinfo {author} {\bibfnamefont {M.}~\bibnamefont {Luskin}}, \bibinfo {author} {\bibfnamefont {E.~B.}\ \bibnamefont {Tadmor}}, \bibinfo {author} {\bibfnamefont {E.}~\bibnamefont {Kaxiras}}, \
  and\ \bibinfo {author} {\bibfnamefont {P.}~\bibnamefont {Kim}},\ }\href {\doibase 10.1038/s41563-019-0346-z} {\bibfield  {journal} {\bibinfo  {journal} {Nature Materials}\ }\textbf {\bibinfo {volume} {18}},\ \bibinfo {pages} {448} (\bibinfo {year} {2019})}\BibitemShut {NoStop}%
\bibitem [{\citenamefont {Ugeda}\ \emph {et~al.}(2014)\citenamefont {Ugeda}, \citenamefont {Bradley}, \citenamefont {Shi}, \citenamefont {Felipe}, \citenamefont {Zhang}, \citenamefont {Qiu}, \citenamefont {Ruan}, \citenamefont {Mo}, \citenamefont {Hussain}, \citenamefont {Shen} \emph {et~al.}}]{ugedaGiantBandgapRenormalization2014}%
  \BibitemOpen
  \bibfield  {author} {\bibinfo {author} {\bibfnamefont {M.~M.}\ \bibnamefont {Ugeda}}, \bibinfo {author} {\bibfnamefont {A.~J.}\ \bibnamefont {Bradley}}, \bibinfo {author} {\bibfnamefont {S.-F.}\ \bibnamefont {Shi}}, \bibinfo {author} {\bibfnamefont {H.}~\bibnamefont {Felipe}}, \bibinfo {author} {\bibfnamefont {Y.}~\bibnamefont {Zhang}}, \bibinfo {author} {\bibfnamefont {D.~Y.}\ \bibnamefont {Qiu}}, \bibinfo {author} {\bibfnamefont {W.}~\bibnamefont {Ruan}}, \bibinfo {author} {\bibfnamefont {S.-K.}\ \bibnamefont {Mo}}, \bibinfo {author} {\bibfnamefont {Z.}~\bibnamefont {Hussain}}, \bibinfo {author} {\bibfnamefont {Z.-X.}\ \bibnamefont {Shen}},  \emph {et~al.},\ }\href {\doibase 10.1038/nmat4061} {\bibfield  {journal} {\bibinfo  {journal} {Nat. Mater.}\ }\textbf {\bibinfo {volume} {13}},\ \bibinfo {pages} {1091} (\bibinfo {year} {2014})}\BibitemShut {NoStop}%
\bibitem [{\citenamefont {Cochrane}\ \emph {et~al.}(2021)\citenamefont {Cochrane}, \citenamefont {Lee}, \citenamefont {Kastl}, \citenamefont {Haber}, \citenamefont {Zhang}, \citenamefont {Kozhakhmetov}, \citenamefont {Robinson}, \citenamefont {Terrones}, \citenamefont {Repp}, \citenamefont {Neaton}, \citenamefont {{Weber-Bargioni}},\ and\ \citenamefont {Schuler}}]{cochraneSpindependentvibronicresponse2021c}%
  \BibitemOpen
  \bibfield  {author} {\bibinfo {author} {\bibfnamefont {K.~A.}\ \bibnamefont {Cochrane}}, \bibinfo {author} {\bibfnamefont {J.-H.}\ \bibnamefont {Lee}}, \bibinfo {author} {\bibfnamefont {C.}~\bibnamefont {Kastl}}, \bibinfo {author} {\bibfnamefont {J.~B.}\ \bibnamefont {Haber}}, \bibinfo {author} {\bibfnamefont {T.}~\bibnamefont {Zhang}}, \bibinfo {author} {\bibfnamefont {A.}~\bibnamefont {Kozhakhmetov}}, \bibinfo {author} {\bibfnamefont {J.~A.}\ \bibnamefont {Robinson}}, \bibinfo {author} {\bibfnamefont {M.}~\bibnamefont {Terrones}}, \bibinfo {author} {\bibfnamefont {J.}~\bibnamefont {Repp}}, \bibinfo {author} {\bibfnamefont {J.~B.}\ \bibnamefont {Neaton}}, \bibinfo {author} {\bibfnamefont {A.}~\bibnamefont {{Weber-Bargioni}}}, \ and\ \bibinfo {author} {\bibfnamefont {B.}~\bibnamefont {Schuler}},\ }\href {\doibase 10.1038/s41467-021-27585-x} {\bibfield  {journal} {\bibinfo  {journal} {Nat. Commun.}\ }\textbf {\bibinfo {volume} {12}},\ \bibinfo {pages} {7287} (\bibinfo {year} {2021})}\BibitemShut {NoStop}%
\end{thebibliography}%


\begin{thebibliography}{10}
\expandafter\ifx\csname url\endcsname\relax
  \def\url#1{\texttt{#1}}\fi
\expandafter\ifx\csname urlprefix\endcsname\relax\def\urlprefix{URL }\fi
\providecommand{\bibinfo}[2]{#2}
\providecommand{\eprint}[2][]{\url{#2}}

\bibitem{aharonovichSolidstatesinglephotonemitters2016b}
\bibinfo{author}{Aharonovich, I.}, \bibinfo{author}{Englund, D.} \& \bibinfo{author}{Toth, M.}
\newblock \href{http://dx.doi.org/10.1038/nphoton.2016.186}{\bibinfo{title}{Solid-state single-photon emitters}}.
\newblock \emph{\bibinfo{journal}{Nat. Photonics}} \textbf{\bibinfo{volume}{10}}, \bibinfo{pages}{631} (\bibinfo{year}{2016}).

\bibitem{atatureMaterialPlatformsSpinbased2018a}
\bibinfo{author}{Atat{\"u}re, M.}, \bibinfo{author}{Englund, D.}, \bibinfo{author}{Vamivakas, N.}, \bibinfo{author}{Lee, S.-Y.} \& \bibinfo{author}{Wrachtrup, J.}
\newblock \bibinfo{title}{Material platforms for spin-based photonic quantum technologies}.
\newblock \emph{\bibinfo{journal}{Nat. Rev. Mater.}} \textbf{\bibinfo{volume}{3}}, \bibinfo{pages}{38--51} (\bibinfo{year}{2018}).

\bibitem{wolfowiczQuantumguidelinessolidstate2021}
\bibinfo{author}{Wolfowicz, G.} \emph{et~al.}
\newblock \href{http://dx.doi.org/10.1038/s41578-021-00306-y}{\bibinfo{title}{Quantum guidelines for solid-state spin defects}}.
\newblock \emph{\bibinfo{journal}{Nat. Rev. Mater.}} \textbf{\bibinfo{volume}{6}}, \bibinfo{pages}{906--925} (\bibinfo{year}{2021}).

\bibitem{gruberScanningconfocaloptical1997}
\bibinfo{author}{Gruber, A.} \emph{et~al.}
\newblock \bibinfo{title}{Scanning confocal optical microscopy and magnetic resonance on single defect centers}.
\newblock \emph{\bibinfo{journal}{Science}} \textbf{\bibinfo{volume}{276}}, \bibinfo{pages}{2012--2014} (\bibinfo{year}{1997}).

\bibitem{castellettosiliconcarbideroomtemperature2014a}
\bibinfo{author}{Castelletto, S.} \emph{et~al.}
\newblock \href{http://dx.doi.org/10.1038/nmat3806}{\bibinfo{title}{A silicon carbide room-temperature single-photon source}}.
\newblock \emph{\bibinfo{journal}{Nature Materials}} \textbf{\bibinfo{volume}{13}}, \bibinfo{pages}{151--156} (\bibinfo{year}{2014}).

\bibitem{maurerRoomtemperaturequantumbit2012a}
\bibinfo{author}{Maurer, P.~C.} \emph{et~al.}
\newblock \bibinfo{title}{Room-temperature quantum bit memory exceeding one second}.
\newblock \emph{\bibinfo{journal}{Science}} \textbf{\bibinfo{volume}{336}}, \bibinfo{pages}{1283--1286} (\bibinfo{year}{2012}).

\bibitem{liu2Dmaterialsquantum2019}
\bibinfo{author}{Liu, X.} \& \bibinfo{author}{Hersam, M.~C.}
\newblock \href{http://dx.doi.org/10.1038/s41578-019-0136-x}{\bibinfo{title}{{{2D}} materials for quantum information science}}.
\newblock \emph{\bibinfo{journal}{Nature Reviews Materials}} \textbf{\bibinfo{volume}{4}}, \bibinfo{pages}{669--684} (\bibinfo{year}{2019}).

\bibitem{moody2022Roadmapintegrated2022}
\bibinfo{author}{Moody, G.} \emph{et~al.}
\newblock \href{http://dx.doi.org/10.1088/2515-7647/ac1ef4}{\bibinfo{title}{2022 {{Roadmap}} on integrated quantum photonics}}.
\newblock \emph{\bibinfo{journal}{Journal of Physics: Photonics}} \textbf{\bibinfo{volume}{4}}, \bibinfo{pages}{012501} (\bibinfo{year}{2022}).

\bibitem{brotons-gisbertQuantumPhotonics2D2023}
\bibinfo{author}{{Brotons-Gisbert}, M.} \& \bibinfo{author}{Gerardot, B.~D.}
\newblock \bibinfo{title}{Quantum {{Photonics}} with {{2D Semiconductors}}}.
\newblock In \emph{\bibinfo{booktitle}{Photonic {{Quantum Technologies}}}}, \bibinfo{pages}{563--579} (\bibinfo{publisher}{John Wiley \& Sons, Ltd}, \bibinfo{year}{2023}).

\bibitem{komsaTwoDimensionalTransitionMetal2012a}
\bibinfo{author}{Komsa, H.-P.} \emph{et~al.}
\newblock \href{http://dx.doi.org/10.1103/PhysRevLett.109.035503}{\bibinfo{title}{Two-{{Dimensional Transition Metal Dichalcogenides}} under {{Electron Irradiation}}: {{Defect Production}} and {{Doping}}}}.
\newblock \emph{\bibinfo{journal}{Physical Review Letters}} \textbf{\bibinfo{volume}{109}}, \bibinfo{pages}{035503} (\bibinfo{year}{2012}).

\bibitem{wangAtomicstructuredefects2018a}
\bibinfo{author}{Wang, S.}, \bibinfo{author}{Robertson, A.} \& \bibinfo{author}{H.~Warner, J.}
\newblock \href{http://dx.doi.org/10.1039/C8CS00236C}{\bibinfo{title}{Atomic structure of defects and dopants in {{2D}} layered transition metal dichalcogenides}}.
\newblock \emph{\bibinfo{journal}{Chemical Society Reviews}} \textbf{\bibinfo{volume}{47}}, \bibinfo{pages}{6764--6794} (\bibinfo{year}{2018}).

\bibitem{barjaIdentifyingSubstitutionalOxygen2019a}
\bibinfo{author}{Barja, S.} \emph{et~al.}
\newblock \bibinfo{title}{Identifying {{Substitutional Oxygen}} as a {{Prolific Point Defect}} in {{Monolayer Transition Metal Dichalcogenides}} with {{Experiment}} and {{Theory}}}.
\newblock \emph{\bibinfo{journal}{Nat. Commun.}} \textbf{\bibinfo{volume}{10}}, \bibinfo{pages}{3382} (\bibinfo{year}{2019}).

\bibitem{schulerHowSubstitutionalPoint2019a}
\bibinfo{author}{Schuler, B.} \emph{et~al.}
\newblock \href{http://dx.doi.org/10.1021/acsnano.9b04611}{\bibinfo{title}{How {{Substitutional Point Defects}} in {{Two-Dimensional WS2 Induce Charge Localization}}, {{Spin}}--{{Orbit Splitting}}, and {{Strain}}}}.
\newblock \emph{\bibinfo{journal}{ACS Nano}} \textbf{\bibinfo{volume}{13}}, \bibinfo{pages}{10520--10534} (\bibinfo{year}{2019}).

\bibitem{schulerLargeSpinOrbitSplitting2019b}
\bibinfo{author}{Schuler, B.} \emph{et~al.}
\newblock \bibinfo{title}{Large {{Spin-Orbit Splitting}} of {{Deep In-Gap Defect States}} of {{Engineered Sulfur Vacancies}} in {{Monolayer WS2}}}.
\newblock \emph{\bibinfo{journal}{Phys. Rev. Lett.}} \textbf{\bibinfo{volume}{123}}, \bibinfo{pages}{076801} (\bibinfo{year}{2019}).

\bibitem{tranQuantumemissionhexagonal2016}
\bibinfo{author}{Tran, T.~T.}, \bibinfo{author}{Bray, K.}, \bibinfo{author}{Ford, M.~J.}, \bibinfo{author}{Toth, M.} \& \bibinfo{author}{Aharonovich, I.}
\newblock \bibinfo{title}{Quantum emission from hexagonal boron nitride monolayers}.
\newblock \emph{\bibinfo{journal}{Nat. Nanotechnol.}} \textbf{\bibinfo{volume}{11}}, \bibinfo{pages}{37} (\bibinfo{year}{2016}).

\bibitem{gottschollInitializationreadoutintrinsic2020}
\bibinfo{author}{Gottscholl, A.} \emph{et~al.}
\newblock \href{http://dx.doi.org/10.1038/s41563-020-0619-6}{\bibinfo{title}{Initialization and read-out of intrinsic spin defects in a van der {{Waals}} crystal at room temperature}}.
\newblock \emph{\bibinfo{journal}{Nature Materials}} \textbf{\bibinfo{volume}{19}}, \bibinfo{pages}{540--545} (\bibinfo{year}{2020}).

\bibitem{hayeeRevealingmultipleclasses2020a}
\bibinfo{author}{Hayee, F.} \emph{et~al.}
\newblock \href{http://dx.doi.org/10.1038/s41563-020-0616-9}{\bibinfo{title}{Revealing multiple classes of stable quantum emitters in hexagonal boron nitride with correlated optical and electron microscopy}}.
\newblock \emph{\bibinfo{journal}{Nature Materials}} \textbf{\bibinfo{volume}{19}}, \bibinfo{pages}{534--539} (\bibinfo{year}{2020}).

\bibitem{mendelsonIdentifyingcarbonsource2020}
\bibinfo{author}{Mendelson, N.} \emph{et~al.}
\newblock \bibinfo{title}{Identifying carbon as the source of visible single-photon emission from hexagonal boron nitride}.
\newblock \emph{\bibinfo{journal}{Nat. Mater.}} \bibinfo{pages}{1--8} (\bibinfo{year}{2020}).

\bibitem{Qiu:2024qyq}
\bibinfo{author}{Qiu, Z.} \emph{et~al.}
\newblock \href{http://dx.doi.org/10.1021/acsnano.4c03640}{\bibinfo{title}{Atomic and {{Electronic Structure}} of {{Defects}} in {{hBN}}: {{Enhancing Single-Defect Functionalities}}}}.
\newblock \emph{\bibinfo{journal}{ACS Nano}} \textbf{\bibinfo{volume}{18}}, \bibinfo{pages}{24035--24043} (\bibinfo{year}{2024}).

\bibitem{Srivastava:2015zvh}
\bibinfo{author}{Srivastava, A.} \emph{et~al.}
\newblock \href{http://dx.doi.org/10.1038/nnano.2015.60}{\bibinfo{title}{Optically active quantum dots in monolayer {{WSe2}}}}.
\newblock \emph{\bibinfo{journal}{Nature Nanotech.}} \textbf{\bibinfo{volume}{10}}, \bibinfo{pages}{491--496} (\bibinfo{year}{2015}).

\bibitem{He:2015pup}
\bibinfo{author}{He, Y.-M.} \emph{et~al.}
\newblock \href{http://dx.doi.org/10.1038/nnano.2015.75}{\bibinfo{title}{Single quantum emitters in monolayer semiconductors}}.
\newblock \emph{\bibinfo{journal}{Nature Nanotech.}} \textbf{\bibinfo{volume}{10}}, \bibinfo{pages}{497--502} (\bibinfo{year}{2015}).

\bibitem{grossoTunablehighpurityroom2017a}
\bibinfo{author}{Grosso, G.} \emph{et~al.}
\newblock \href{http://dx.doi.org/10.1038/s41467-017-00810-2}{\bibinfo{title}{Tunable and high-purity room temperature single-photon emission from atomic defects in hexagonal boron nitride}}.
\newblock \emph{\bibinfo{journal}{Nature Communications}} \textbf{\bibinfo{volume}{8}}, \bibinfo{pages}{705} (\bibinfo{year}{2017}).

\bibitem{Parto:2020fmx}
\bibinfo{author}{Parto, K.}, \bibinfo{author}{Azzam, S.~I.}, \bibinfo{author}{Banerjee, K.} \& \bibinfo{author}{Moody, G.}
\newblock \href{http://dx.doi.org/10.1038/s41467-021-23709-5}{\bibinfo{title}{Defect and strain engineering of monolayer {{WSe2}} enables site-controlled single-photon emission up to 150 {{K}}}}.
\newblock \emph{\bibinfo{journal}{Nature Commun.}} \textbf{\bibinfo{volume}{12}}, \bibinfo{pages}{3585} (\bibinfo{year}{2021}).

\bibitem{kleinSiteselectivelygeneratedphoton2019a}
\bibinfo{author}{Klein, J.} \emph{et~al.}
\newblock \href{http://dx.doi.org/10.1038/s41467-019-10632-z}{\bibinfo{title}{Site-selectively generated photon emitters in monolayer {{MoS2}} via local helium ion irradiation}}.
\newblock \emph{\bibinfo{journal}{Nature Communications}} \textbf{\bibinfo{volume}{10}}, \bibinfo{pages}{2755} (\bibinfo{year}{2019}).

\bibitem{sunUnveilingsulfurvacancy2024}
\bibinfo{author}{Sun, H.} \emph{et~al.}
\newblock \href{http://dx.doi.org/10.1038/s41467-024-53880-4}{\bibinfo{title}{Unveiling sulfur vacancy pairs as bright and stable color centers in monolayer {{WS2}}}}.
\newblock \emph{\bibinfo{journal}{Nature Communications}} \textbf{\bibinfo{volume}{15}}, \bibinfo{pages}{9476} (\bibinfo{year}{2024}).

\bibitem{baoVisualizingNanoscaleExcitonic2015}
\bibinfo{author}{Bao, W.} \emph{et~al.}
\newblock \href{http://dx.doi.org/10.1038/ncomms8993}{\bibinfo{title}{Visualizing nanoscale excitonic relaxation properties of disordered edges and grain boundaries in monolayer molybdenum disulfide}}.
\newblock \emph{\bibinfo{journal}{Nature Communications}} \textbf{\bibinfo{volume}{6}}, \bibinfo{pages}{7993} (\bibinfo{year}{2015}).

\bibitem{zhangNanospectroscopyExcitonsAtomically2022}
\bibinfo{author}{Zhang, S.} \emph{et~al.}
\newblock \href{http://dx.doi.org/10.1038/s41467-022-28117-x}{\bibinfo{title}{Nano-spectroscopy of excitons in atomically thin transition metal dichalcogenides}}.
\newblock \emph{\bibinfo{journal}{Nature Communications}} \textbf{\bibinfo{volume}{13}}, \bibinfo{pages}{542} (\bibinfo{year}{2022}).

\bibitem{lohNbimpurityboundexcitons2024}
\bibinfo{author}{Loh, L.} \emph{et~al.}
\newblock \href{http://dx.doi.org/10.1038/s41467-024-54360-5}{\bibinfo{title}{Nb impurity-bound excitons as quantum emitters in monolayer {{WS2}}}}.
\newblock \emph{\bibinfo{journal}{Nature Communications}} \textbf{\bibinfo{volume}{15}}, \bibinfo{pages}{10035} (\bibinfo{year}{2024}).

\bibitem{mitterreiterrolechalcogenvacancies2021}
\bibinfo{author}{Mitterreiter, E.} \emph{et~al.}
\newblock \bibinfo{title}{The role of chalcogen vacancies for atomic defect emission in {{MoS2}}}.
\newblock \emph{\bibinfo{journal}{Nat. Commun.}} \textbf{\bibinfo{volume}{12}}, \bibinfo{pages}{3822} (\bibinfo{year}{2021}).

\bibitem{ugedaGiantbandgaprenormalization2014}
\bibinfo{author}{Ugeda, M.~M.} \emph{et~al.}
\newblock \href{http://dx.doi.org/10.1038/nmat4061}{\bibinfo{title}{Giant bandgap renormalization and excitonic effects in a monolayer transition metal dichalcogenide semiconductor}}.
\newblock \emph{\bibinfo{journal}{Nat. Mater.}} \textbf{\bibinfo{volume}{13}}, \bibinfo{pages}{1091} (\bibinfo{year}{2014}).

\bibitem{lorchatFilteringphotoluminescencespectra2020a}
\bibinfo{author}{Lorchat, E.} \emph{et~al.}
\newblock \href{http://dx.doi.org/10.1038/s41565-020-0644-2}{\bibinfo{title}{Filtering the photoluminescence spectra of atomically thin semiconductors with graphene}}.
\newblock \emph{\bibinfo{journal}{Nat. Nanotechnol.}} \textbf{\bibinfo{volume}{15}}, \bibinfo{pages}{283--288} (\bibinfo{year}{2020}).

\bibitem{nisiScanningprobespectroscopy2024}
\bibinfo{author}{Nisi, K.} \emph{et~al.}
\newblock \href{http://dx.doi.org/10.1088/2053-1583/ada046}{\bibinfo{title}{Scanning probe spectroscopy of sulfur vacancies and {{MoS2}} monolayers in side-contacted van der {{Waals}} heterostructures}}.
\newblock \emph{\bibinfo{journal}{2D Materials}} \textbf{\bibinfo{volume}{12}}, \bibinfo{pages}{015023} (\bibinfo{year}{2024}).

\bibitem{doppagneElectrofluorochromismsinglemoleculelevel2018}
\bibinfo{author}{Doppagne, B.} \emph{et~al.}
\newblock \bibinfo{title}{Electrofluorochromism at the single-molecule level}.
\newblock \emph{\bibinfo{journal}{Science}} \textbf{\bibinfo{volume}{361}}, \bibinfo{pages}{251--255} (\bibinfo{year}{2018}).

\bibitem{miwaManyBodyStateDescription2019a}
\bibinfo{author}{Miwa, K.} \emph{et~al.}
\newblock \href{http://dx.doi.org/10.1021/acs.nanolett.8b04484}{\bibinfo{title}{Many-{{Body State Description}} of {{Single-Molecule Electroluminescence Driven}} by a {{Scanning Tunneling Microscope}}}}.
\newblock \emph{\bibinfo{journal}{Nano Letters}} \textbf{\bibinfo{volume}{19}}, \bibinfo{pages}{2803--2811} (\bibinfo{year}{2019}).

\bibitem{jiangManyBodyDescriptionSTMInduced2023a}
\bibinfo{author}{Jiang, S.} \emph{et~al.}
\newblock \href{http://dx.doi.org/10.1103/PhysRevLett.130.126202}{\bibinfo{title}{Many-{{Body Description}} of {{STM-Induced Fluorescence}} of {{Charged Molecules}}}}.
\newblock \emph{\bibinfo{journal}{Physical Review Letters}} \textbf{\bibinfo{volume}{130}}, \bibinfo{pages}{126202} (\bibinfo{year}{2023}).

\bibitem{cochraneIntentionalcarbondoping2020b}
\bibinfo{author}{Cochrane, K.~A.} \emph{et~al.}
\newblock \bibinfo{title}{Intentional carbon doping reveals {{CH}} as an abundant charged impurity in nominally undoped synthetic {{WS}}{\textsubscript{2}} and {{WSe}}{\textsubscript{2}}}.
\newblock \emph{\bibinfo{journal}{2D Mater.}} \textbf{\bibinfo{volume}{7}}, \bibinfo{pages}{031003} (\bibinfo{year}{2020}).

\bibitem{riis-jensenElectricallycontrolleddielectric2020}
\bibinfo{author}{{Riis-Jensen}, A.~C.}, \bibinfo{author}{Lu, J.} \& \bibinfo{author}{Thygesen, K.~S.}
\newblock \href{http://dx.doi.org/10.1103/PhysRevB.101.121110}{\bibinfo{title}{Electrically controlled dielectric band gap engineering in a two-dimensional semiconductor}}.
\newblock \emph{\bibinfo{journal}{Physical Review B}} \textbf{\bibinfo{volume}{101}}, \bibinfo{pages}{121110} (\bibinfo{year}{2020}).

\bibitem{mitterreiterAtomisticPositioningDefects2020a}
\bibinfo{author}{Mitterreiter, E.} \emph{et~al.}
\newblock \bibinfo{title}{Atomistic {{Positioning}} of {{Defects}} in {{Helium Ion Treated Single-Layer MoS2}}}.
\newblock \emph{\bibinfo{journal}{Nano Lett.}} \textbf{\bibinfo{volume}{20}}, \bibinfo{pages}{4437--4444} (\bibinfo{year}{2020}).

\bibitem{xiangChargestatedependentsymmetry2024a}
\bibinfo{author}{Xiang, F.} \emph{et~al.}
\newblock \href{http://dx.doi.org/10.1038/s41467-024-47039-4}{\bibinfo{title}{Charge state-dependent symmetry breaking of atomic defects in transition metal dichalcogenides}}.
\newblock \emph{\bibinfo{journal}{Nature Communications}} \textbf{\bibinfo{volume}{15}}, \bibinfo{pages}{2738} (\bibinfo{year}{2024}).

\bibitem{bobzienLayerDependentChargeStateLifetime2025a}
\bibinfo{author}{Bobzien, L.} \emph{et~al.}
\newblock \href{http://dx.doi.org/10.1103/PhysRevLett.134.076201}{\bibinfo{title}{Layer-{{Dependent Charge-State Lifetime}} of {{Single Se Vacancies}} in {{WSe2}}}}.
\newblock \emph{\bibinfo{journal}{Physical Review Letters}} \textbf{\bibinfo{volume}{134}}, \bibinfo{pages}{076201} (\bibinfo{year}{2025}).

\bibitem{withersLightemittingdiodesbandstructure2015a}
\bibinfo{author}{Withers, F.} \emph{et~al.}
\newblock \href{http://dx.doi.org/10.1038/nmat4205}{\bibinfo{title}{Light-emitting diodes by band-structure engineering in van der {{Waals}} heterostructures}}.
\newblock \emph{\bibinfo{journal}{Nature Materials}} \textbf{\bibinfo{volume}{14}}, \bibinfo{pages}{301--306} (\bibinfo{year}{2015}).

\bibitem{palacios-berraqueroAtomicallythinquantum2016}
\bibinfo{author}{{Palacios-Berraquero}, C.} \emph{et~al.}
\newblock \href{http://dx.doi.org/10.1038/ncomms12978}{\bibinfo{title}{Atomically thin quantum light-emitting diodes}}.
\newblock \emph{\bibinfo{journal}{Nature Communications}} \textbf{\bibinfo{volume}{7}}, \bibinfo{pages}{12978} (\bibinfo{year}{2016}).

\bibitem{wangExcitonassistedelectrontunnelling2023}
\bibinfo{author}{Wang, L.} \emph{et~al.}
\newblock \href{http://dx.doi.org/10.1038/s41563-023-01556-7}{\bibinfo{title}{Exciton-assisted electron tunnelling in van der {{Waals}} heterostructures}}.
\newblock \emph{\bibinfo{journal}{Nature Materials}} \textbf{\bibinfo{volume}{22}}, \bibinfo{pages}{1094--1099} (\bibinfo{year}{2023}).

\bibitem{pommierScanningTunnelingMicroscopeInduced2019}
\bibinfo{author}{Pommier, D.} \emph{et~al.}
\newblock \href{http://dx.doi.org/10.1103/PhysRevLett.123.027402}{\bibinfo{title}{Scanning {{Tunneling Microscope-Induced Excitonic Luminescence}} of a {{Two-Dimensional Semiconductor}}}}.
\newblock \emph{\bibinfo{journal}{Phys. Rev. Lett.}} \textbf{\bibinfo{volume}{123}}, \bibinfo{pages}{027402} (\bibinfo{year}{2019}).

\bibitem{schulerElectricallydrivenphoton2020a}
\bibinfo{author}{Schuler, B.} \emph{et~al.}
\newblock \bibinfo{title}{Electrically driven photon emission from individual atomic defects in monolayer {{WS2}}}.
\newblock \emph{\bibinfo{journal}{Sci. Adv.}} \textbf{\bibinfo{volume}{6}}, \bibinfo{pages}{eabb5988} (\bibinfo{year}{2020}).

\bibitem{romanTunnelingcurrentinducedlocalexcitonic2020}
\bibinfo{author}{Rom{\'a}n, R. J.~P.} \emph{et~al.}
\newblock \href{http://dx.doi.org/10.1039/D0NR03400B}{\bibinfo{title}{Tunneling-current-induced local excitonic luminescence in p-doped {{WSe2}} monolayers}}.
\newblock \emph{\bibinfo{journal}{Nanoscale}} \textbf{\bibinfo{volume}{12}}, \bibinfo{pages}{13460--13470} (\bibinfo{year}{2020}).

\bibitem{lopezTipinducedexcitonicluminescence2023}
\bibinfo{author}{L{\'o}pez, L. E.~P.}, \bibinfo{author}{Ros{\l}awska, A.}, \bibinfo{author}{Scheurer, F.}, \bibinfo{author}{Berciaud, S.} \& \bibinfo{author}{Schull, G.}
\newblock \href{http://dx.doi.org/10.1038/s41563-023-01494-4}{\bibinfo{title}{Tip-induced excitonic luminescence nanoscopy of an atomically resolved van der {{Waals}} heterostructure}}.
\newblock \emph{\bibinfo{journal}{Nature Materials}} \bibinfo{pages}{1--7} (\bibinfo{year}{2023}).

\bibitem{gengImagingValleyExcitons2024}
\bibinfo{author}{Geng, H.} \emph{et~al.}
\newblock \href{http://dx.doi.org/10.1021/acsnano.3c12555}{\bibinfo{title}{Imaging {{Valley Excitons}} in a {{2D Semiconductor}} with {{Scanning Tunneling Microscope-Induced Luminescence}}}}.
\newblock \emph{\bibinfo{journal}{ACS Nano}} \textbf{\bibinfo{volume}{18}}, \bibinfo{pages}{8961--8970} (\bibinfo{year}{2024}).

\bibitem{qianProbingDarkExcitons2024}
\bibinfo{author}{Qian, C.} \emph{et~al.}
\newblock \href{http://dx.doi.org/10.1103/PhysRevLett.133.086902}{\bibinfo{title}{Probing {{Dark Excitons}} in {{Monolayer MoS2}} by {{Nonlinear Two-Photon Spectroscopy}}}}.
\newblock \emph{\bibinfo{journal}{Physical Review Letters}} \textbf{\bibinfo{volume}{133}}, \bibinfo{pages}{086902} (\bibinfo{year}{2024}).

\bibitem{druppelDiversitytrionstates2017}
\bibinfo{author}{Dr{\"u}ppel, M.}, \bibinfo{author}{Deilmann, T.}, \bibinfo{author}{Kr{\"u}ger, P.} \& \bibinfo{author}{Rohlfing, M.}
\newblock \href{http://dx.doi.org/10.1038/s41467-017-02286-6}{\bibinfo{title}{Diversity of trion states and substrate effects in the optical properties of an {{MoS2}} monolayer}}.
\newblock \emph{\bibinfo{journal}{Nature Communications}} \textbf{\bibinfo{volume}{8}}, \bibinfo{pages}{2117} (\bibinfo{year}{2017}).

\bibitem{grzeszczykExposingtrionsfine2021}
\bibinfo{author}{Grzeszczyk, M.} \emph{et~al.}
\newblock \href{http://dx.doi.org/10.1039/D1NR03855A}{\bibinfo{title}{Exposing the trion's fine structure by controlling the carrier concentration in {{hBN-encapsulated MoS2}}}}.
\newblock \emph{\bibinfo{journal}{Nanoscale}} \textbf{\bibinfo{volume}{13}}, \bibinfo{pages}{18726--18733} (\bibinfo{year}{2021}).

\bibitem{kleinTrions$mathrmMoS_2$are2022}
\bibinfo{author}{Klein, J.} \emph{et~al.}
\newblock \href{http://dx.doi.org/10.1103/PhysRevB.105.L041302}{\bibinfo{title}{Trions in {{MoS2}} are quantum superpositions of intra- and intervalley spin states}}.
\newblock \emph{\bibinfo{journal}{Physical Review B}} \textbf{\bibinfo{volume}{105}}, \bibinfo{pages}{L041302} (\bibinfo{year}{2022}).

\bibitem{aghajanianResonantboundstates2020}
\bibinfo{author}{Aghajanian, M.} \emph{et~al.}
\newblock \bibinfo{title}{Resonant and bound states of charged defects in two-dimensional semiconductors}.
\newblock \emph{\bibinfo{journal}{Physical Review B}} \textbf{\bibinfo{volume}{101}}, \bibinfo{pages}{081201} (\bibinfo{year}{2020}).

\bibitem{cochraneSpindependentvibronicresponse2021c}
\bibinfo{author}{Cochrane, K.~A.} \emph{et~al.}
\newblock \href{http://dx.doi.org/10.1038/s41467-021-27585-x}{\bibinfo{title}{Spin-dependent vibronic response of a carbon radical ion in two-dimensional {{WS2}}}}.
\newblock \emph{\bibinfo{journal}{Nat. Commun.}} \textbf{\bibinfo{volume}{12}}, \bibinfo{pages}{7287} (\bibinfo{year}{2021}).

\bibitem{chernikovExcitonBindingEnergy2014a}
\bibinfo{author}{Chernikov, A.} \emph{et~al.}
\newblock \href{http://dx.doi.org/10.1103/PhysRevLett.113.076802}{\bibinfo{title}{Exciton {{Binding Energy}} and {{Nonhydrogenic Rydberg Series}} in {{Monolayer WS2}}}}.
\newblock \emph{\bibinfo{journal}{Physical Review Letters}} \textbf{\bibinfo{volume}{113}}, \bibinfo{pages}{076802} (\bibinfo{year}{2014}).

\bibitem{wangColloquiumExcitonsAtomically2018b}
\bibinfo{author}{Wang, G.} \emph{et~al.}
\newblock \href{http://dx.doi.org/10.1103/RevModPhys.90.021001}{\bibinfo{title}{Colloquium: {{Excitons}} in atomically thin transition metal dichalcogenides}}.
\newblock \emph{\bibinfo{journal}{Reviews of Modern Physics}} \textbf{\bibinfo{volume}{90}}, \bibinfo{pages}{021001} (\bibinfo{year}{2018}).

\bibitem{yeProbingexcitonicdark2014}
\bibinfo{author}{Ye, Z.} \emph{et~al.}
\newblock \href{http://dx.doi.org/10.1038/nature13734}{\bibinfo{title}{Probing excitonic dark states in single-layer tungsten disulphide}}.
\newblock \emph{\bibinfo{journal}{Nature}} \textbf{\bibinfo{volume}{513}}, \bibinfo{pages}{214--218} (\bibinfo{year}{2014}).

\bibitem{pankoveOpticalProcessesSemiconductors2012}
\bibinfo{author}{Pankove, J.~I.}
\newblock \bibinfo{title}{Optical {{Processes}} in {{Semiconductors}}} (\bibinfo{publisher}{Courier Corporation}, \bibinfo{year}{2012}).

\bibitem{mostaaniDiffusionquantumMonte2017a}
\bibinfo{author}{Mostaani, E.} \emph{et~al.}
\newblock \href{http://dx.doi.org/10.1103/PhysRevB.96.075431}{\bibinfo{title}{Diffusion quantum {{Monte Carlo}} study of excitonic complexes in two-dimensional transition-metal dichalcogenides}}.
\newblock \emph{\bibinfo{journal}{Physical Review B}} \textbf{\bibinfo{volume}{96}}, \bibinfo{pages}{075431} (\bibinfo{year}{2017}).

\end{thebibliography}
\clearpage

\section*{Figures}

\clearpage

\clearpage

\clearpage

\clearpage

\section*{Data availability}
The data that support the findings of this study are available from the corresponding author upon request.

\section*{Acknowledgements}
The authors thank Roman Fasel, Nils Krane, Tom\'a\v{s} Neuman, Christoph Kastl, and Adam Gali for fruitful discussions.
This research was funded by the European Research Council (ERC) under the European Union's Horizon 2020 research and innovation program (Grant agreement No. 948243). SEA, NK and OG appreciate financial support from the Werner Siemens Foundation (CarboQuant).
S.W. acknowledges financial support from the National Natural Science Foundation of China (Grants No. 22325203 and No. 92365302) and the Shanghai Jiao Tong University 2030 Initiative.
L.N. and S.P. acknowledge support by the Swiss National Science Foundation (grant 200020\_192362/1)
K.W. and T.T. acknowledge support from the JSPS KAKENHI (Grant Numbers 21H05233 and 23H02052) , the CREST (JPMJCR24A5), JST and World Premier International Research Center Initiative (WPI), MEXT, Japan.
J.A.R and C.D. acknowledge the support by 2DCC-MIP under NSF cooperative agreement DMR-2039351 and the Penn State MRSEC Center for Nanoscale Science via NSF award DMR2011839.
For the purpose of Open Access, the author has applied a CC BY public copyright license to any Author Accepted Manuscript version arising from this submission.

\section*{Declarations}
The authors declare no conflict of interest.

\end{document}


\title[SI: MoS2 Exciton on hBN]{Supplemental Material: Atomically-resolved exciton emission from single defects in MoS\textsubscript{2}}

\author{Lysander Huberich}
\affiliation{nanotech@surfaces Laboratory, Empa -- Swiss Federal Laboratories for Materials Science and Technology, D\"ubendorf 8600, Switzerland}

\author{Eve Ammerman}
\affiliation{nanotech@surfaces Laboratory, Empa -- Swiss Federal Laboratories for Materials Science and Technology, D\"ubendorf 8600, Switzerland}

\author{Gu Yu}
\affiliation{Key Laboratory of Artificial Structures and Quantum Control (Ministry of Education), Tsung-Dao Lee Institute, School of Physics and Astronomy, Shanghai Jiao Tong University, Shanghai, 200240, China}

\author{Yining Ren}
\affiliation{Key Laboratory of Artificial Structures and Quantum Control (Ministry of Education), Tsung-Dao Lee Institute, School of Physics and Astronomy, Shanghai Jiao Tong University, Shanghai, 200240, China}

\author{Sotirios Papadopoulos\,\orcidlink{0000-0002-3225-8239}}
\affiliation{Photonics Laboratory, ETH Zürich, Zürich 8093, Switzerland}
\affiliation{Universit\'e de  Strasbourg, CNRS, Institut de Physique et Chimie des  Mat\'erieux de Strasbourg, UMR 7504, F-67000 Strasbourg, France}

\author{Chengye Dong}
\affiliation{Two-Dimensional Crystal Consortium, The Pennsylvania State University, University Park, PA 16802, USA}

\author{Joshua A. Robinson\,\orcidlink{0000-0002-1513-7187}}
\affiliation{Department of Materials Science and Engineering, The Pennsylvania State University, University Park, PA 16082, USA}
\affiliation{Two-Dimensional Crystal Consortium, The Pennsylvania State University, University Park, PA 16802, USA}
\affiliation{Department of Chemistry and Department of Physics, The Pennsylvania State University, University Park, PA, 16802, USA}

\author{Kenji Watanabe\,\orcidlink{0000-0003-3701-8119}}
\affiliation{Research Center for Electronic and Optical Materials, National Institute for Materials Science, 1-1 Namiki, Tsukuba 305-0044, Japan}

\author{Takashi Taniguchi\,\orcidlink{0000-0002-1467-3105}}
\affiliation{Research Center for Materials Nanoarchitectonics, National Institute for Materials Science,  1-1 Namiki, Tsukuba 305-0044, Japan}

\author{Oliver Gröning}
\affiliation{nanotech@surfaces Laboratory, Empa -- Swiss Federal Laboratories for Materials Science and Technology, D\"ubendorf 8600, Switzerland}

\author{Lukas Novotny}
\affiliation{Photonics Laboratory, ETH Zürich, Zürich 8093, Switzerland}

\author{Tingxin Li\,\orcidlink{0000-0002-3572-4530}}
\affiliation{Key Laboratory of Artificial Structures and Quantum Control (Ministry of Education), Tsung-Dao Lee Institute, School of Physics and Astronomy, Shanghai Jiao Tong University, Shanghai, 200240, China}

\author{Shiyong Wang\,\orcidlink{0000-0001-6603-9926}}
\affiliation{Key Laboratory of Artificial Structures and Quantum Control (Ministry of Education), Tsung-Dao Lee Institute, School of Physics and Astronomy, Shanghai Jiao Tong University, Shanghai, 200240, China}

\author{Bruno Schuler\,\orcidlink{0000-0002-9641-0340}}
\email[]{bruno.schuler@empa.ch}
\affiliation{nanotech@surfaces Laboratory, Empa -- Swiss Federal Laboratories for Materials Science and Technology, D\"ubendorf 8600, Switzerland}

\date{\today}
\pacs{}
\maketitle

\tableofcontents
\newpage

\clearpage

\section*{Methods}

\subsection*{Sample Fabrication and Preparation}
Mechanically exfoliated hBN (National Institute for Materials Science, Japan) and MoS$_2$ (HQGraphene) flakes were characterized optically and by AFM, before transfer by a PC/PDMS stamp ~\cite{zomerFastPickTechnique2014} to either a quasi-freestanding epitaxial graphene on SiC (Pennsylvania State University, USA) or few-layer graphene on hBN/SiO$_2$/Si substrate as detailed in SI Fig.~\ref{fig:SI_Device}. Before measurement, the samples were annealed in UHV for several hours at 350\,$^\circ$C. In order to induce Vac$_\text{S}^-$ and CH$_\text{S}^-$ defects, we mildly Ar-sputtered (1\,s, $3.6\times10^{-6}$\,mbar, 0.2\,kV bias, sub-$\mu$A sputter current)~\cite{maControlledArgonBeaminduced2013b} the sample at room-temperature.

\subsection*{Scanning probe (SPM) and STM-induced luminescence (STML) measurements}
\begin{figure*}[h]
\includegraphics[width=\textwidth]{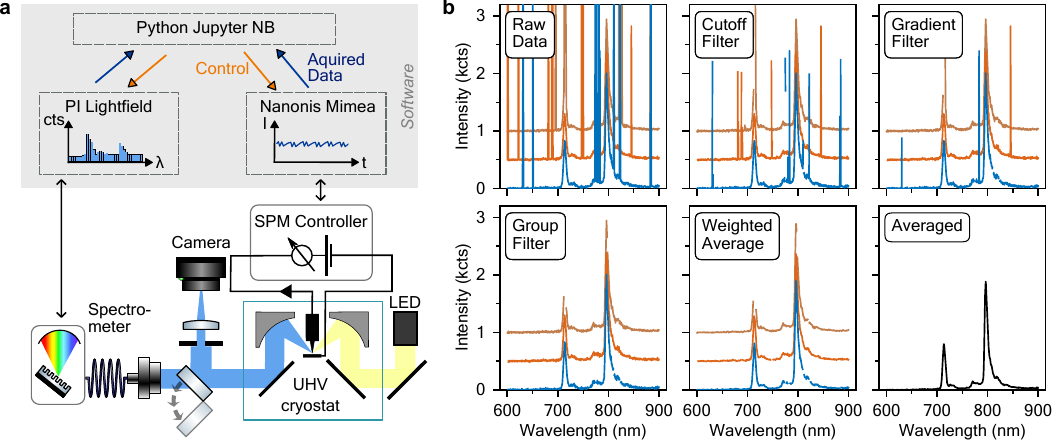}
\caption{\label{fig:SI_STML_Acq_Proc}
\textbf{Schematic Illustration of the STML Setup and Data Processing.}
\textbf{a}, The lower part shows a sketch of the experimental setup used to conduct the (STML) experiments. Tow off-axis parabolic mirrors (NA = 0.4) are used to collect light in STML measurements and visual camera access for the sample approach. Outside UHV, a fiber collimator is used to couple via a fiber into the spectrometer. The upper part depicts the software acquisition procedure. While the optical spectra is recorded via the Princton Instruments Lightfield software, time-traces of essential STM parameters are logged by the Nanonis Mimea software in parallel. Finally all acquired data is written into a single spectra file by a controlling Python Jupyter notebook. 
\textbf{b}, Example of multiple data processing steps used to remove cosmic rays from collected STML spectra. Due to the long integration time and the high sensitivity of the detector, cosmic rays are frequently observed in the recorded optical spectrum. A series of filters are used to remove them: a cutoff filter, a gradient threshold filter and a majority group filter for spectra recorded in triplicate. Finally smoothing is performed when necessary by a local triangular weighted average on each individual spectrum and an average is calculated for triplicate measurements.
}
\end{figure*}

\begin{figure*}[h]
\includegraphics[width=\textwidth]{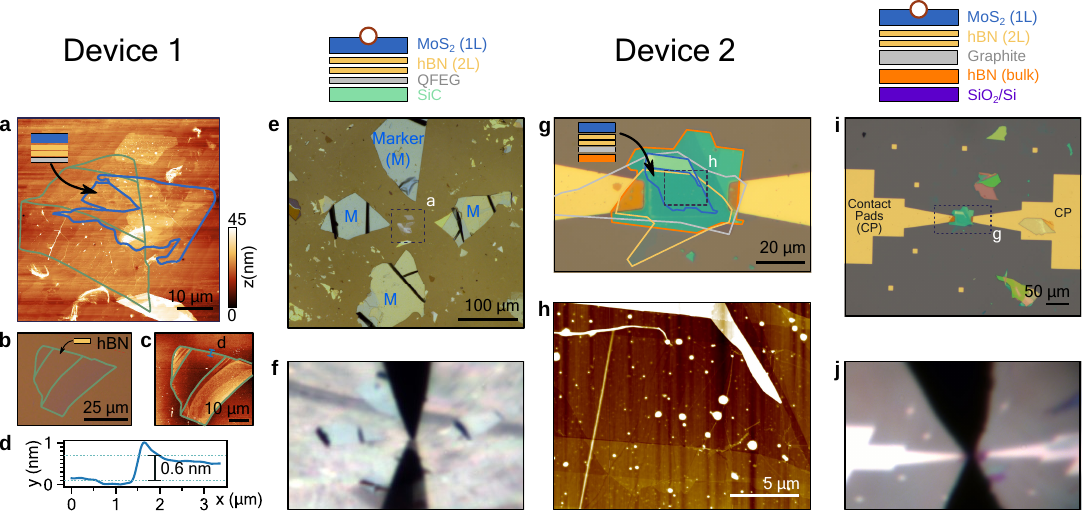}
\caption{\label{fig:SI_Device}
\textbf{Heterostructure device fabrication and optical approach in the STM.}
\textbf{a}, Device 1 (D1) AFM topography after transfer onto QFEG substrate with outlines for the hBN (green) and MoS$_2$ (blue) flakes.
\textbf{b-d}, Optical image (b), AFM (c) and height profile (d) of a 2L hBN flake on 90\,nm SiO$_2$/Si before PC/PDMS-assisted transfer.
\textbf{e}, Optical overview image (D1) with marker flakes. Location of the heterostructure shown in (a) is marked by the dashed box.
\textbf{f}, Optical image (D1) of STM tip positioned over the stack using marker flakes and white-light illumination.
\textbf{g}, Device 2 (D2) optical image after transfer onto pre-pattered SiO$_2$/Si with outlines for the MoS\textsubscript{2} (blue), hBN (2L) (yellow), graphite (grey) and bulk hBN (orange) flakes. The Ti/Pt contacts serve both as navigation pads and sample contacts. AFM image location show in (h) is marked by the dashed box.
\textbf{h}, Close-up AFM image of heterostructure (D2) after fabrication.
\textbf{i}, Optical overview image (D2). Location of the heterostructure shown in (g) is marked by the dashed box.
\textbf{j}, Optical image (D2) of STM tip positioned over the stack. 
}
\end{figure*}

SPM measurements were performed using a CreaTec Fischer \& Co. GmbH scanning probe microscopy at liquid helium temperatures (\(T < 5 \, \text{K}\)) under ultrahigh vacuum (\(p < 2 \times 10^{-10} \, \text{mbar}\)). Optical access to the sample was achieved through parabolic mirrors with a numerical aperture of 0.4. A telescopic camera setup was used for coarse positioning of the tip on the sample heterostructure using marker flakes or contact pads as visual guides. Optical spectra were collected through a fiber collimator using a Teledyne Princeton Instruments SpectraPro HRS 300 spectrograph, equipped with an LN-cooled PyLoN 400BR eXcelon detector. The optical collection setup is further detailed in \figref{fig:SI_STML_Acq_Proc}. PL measurements were conducted \textit{in-situ} using a PicoQuant Prima-205-515-640 laser.\\

Both tungsten tips conditioned on an Au(111) surface and Ag tips conditioned on a Ag(111) surface where used for the measurements. The tip plasmon was optimized for cavity strength and plasmonic peak wavelength, spectrally overlapping with MoS$_2$ and defect emission (see Fig.~\ref{fig:SI_Plasmon}). STM topographic measurements were conducted in constant-current mode, with the bias voltage applied to the sample. STS data were collected using a lock-in amplifier (Nanonis Specs) operating at a resonance frequency between 600 and 900\,Hz.
STML measurements were acquired using an integrated software platform that enabled simultaneous recording of optical and SPM data. All STML spectra in the main text (except Fig. 4c-i) were obtained with a five minute (15 seconds) exposure time per measurement, recorded in triplicate (single), and filtered for cosmic rays as detailed in SI Fig.~\ref{fig:SI_STML_Acq_Proc}. Peak fitting of the STML spectra was performed following the procedure outlined in Fig.~\ref{fig:SI_STML_Fitting}.

\clearpage

\section*{Extended Data figures}

\subsection*{Schematic illustration of STML setup and data processing}

\newpage
\subsection*{Device fabrication and optical approach in the STM}

\newpage
\subsection*{Identifying graphene, hBN and MoS$_2$ by STM and STS}
\begin{figure*}[h]
\includegraphics[width=\textwidth]{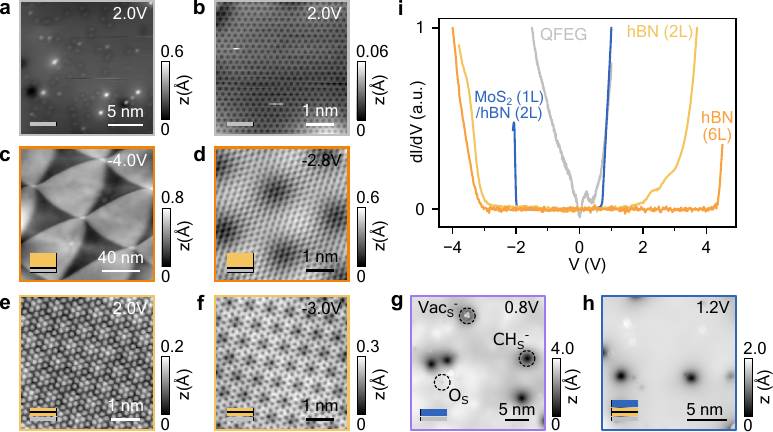}
\caption{\label{fig:SI_Identifying_Areas}
\textbf{Identifying graphene, hBN and MoS$_2$ by STM topography images and STS.}
\textbf{a,b}, STM topography of QFEG at 2\,V.
\textbf{c}, STM topography ($V$ = -4\,V) of bulk ($>$ 5L) hBN showing strain relief patterns from unintentional twisting ~\cite{kazmierczakStrainFieldsTwisted2021,yooAtomicElectronicReconstruction2019}. 
\textbf{d}, STM topography ($V$ = -2.8\,V) of bulk hBN with moiré patterns.
\textbf{e,f}, STM topography of bilayer hBN on QFEG/SiC at 2\,V and -3\,V, respectively.
\textbf{g}, STM image of synthetically grown (MOCVD) monolayer MoS$_2$/QFEG. Common point defects are labelled.
\textbf{h}, STM image of exfoliated monolayer MoS$_2$/hBN(2L)/QFEG.
\textbf{i}, Comparison of d$I$/d$V$ spectra on QFEG (gray), hBN(2L)/QFEG (light yellow), hBN(6L) (1.9\,nm)/QFEG (yellow), and MoS$_2$/hBN(2L)/QFEG (blue).
}
\end{figure*}

\newpage
\subsection*{Layer-dependent STS of hBN/QFEG and moiré pattern formation in hBN(2L)/QFEG}
\begin{figure*} [h]
\includegraphics[width=\textwidth]{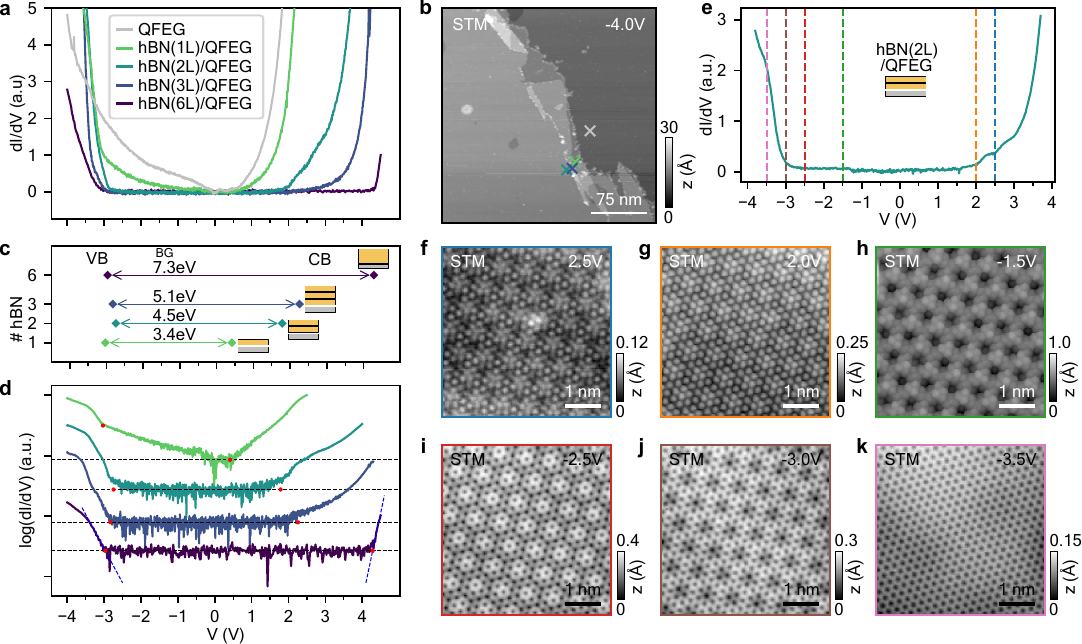}
\caption{\label{fig:SI_hBN_STS}
\textbf{Layer-dependent STS of hBN/QFEG and moiré pattern in hBN(2L)/QFEG}
\textbf{a}, STS of QFEG and hBN/QFEG with varying hBN thicknesses.
\textbf{b}, STM topography of 1-3L hBN flake on QFEG with STS locations from (a) indicated. 
\textbf{c}, Estimated band edges (VB,CB) and bandgap (BG) as extracted from STS shown in (a) without correction for the voltage drop across the double barrier tunneling junction~\cite{ugedaGiantBandgapRenormalization2014}.
\textbf{d}, Vertically offset log-scale STS spectra from (a), with noise floor (dashed) and estimated band edges indicated (red dots).
\textbf{e}, STS of hBN(2L)/QFEG with bias of STM images in (f) indicated as dashed lines.
\textbf{f-k}, STM topography of moiré patterns in hBN(2L)/QFEG at different sample bias.
}
\end{figure*}

\newpage
\subsection*{Electronic band changes in MoS$_{\text{2}}$ due to hBN decoupling}
\begin{figure*} [h]
\includegraphics[width=\textwidth]{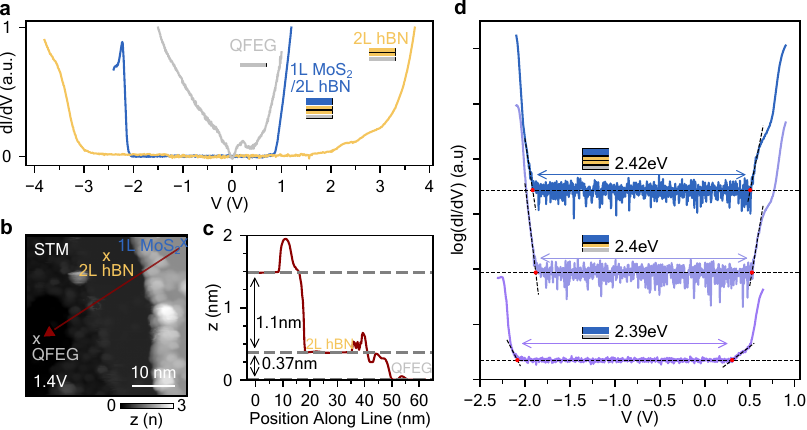}
\caption{\label{fig:SI_MoS2_STS}
\textbf{Electronic band alignment change in MoS$_{\text{2}}$ due to hBN decoupling.}
\textbf{a}, STS spectra acquired on different layers of the heterostructure: QFEG (gray), hBN(2L)/QFEG (yellow), and MoS$_2$(1L)/hBN(2L)/QFEG (blue). Acquisition locations indicated in (b).
\textbf{b}, STM topography acquired at the edge of the heterostructure showing distinct material regions. The red arrow marks the line along which the height profile in (c) was extracted.
\textbf{c}, Topographic height profile extracted from (b), revealing step heights corresponding to sequential stacking of QFEG, hBN(2L) and MoS$_2$. 
\textbf{d}, Vertically offset log-scale plot of STS spectra MoS$_{\text{2}}$(1L)/hBN(0,1,2L)/QFEG with estimated VB and CB onsets, and bandgap values (BG)~\cite{ugedaGiantBandgapRenormalization2014}. The values are not corrected for the voltage drop across the double barrier tunneling junction.
}
\end{figure*}

\newpage
\subsection*{Band onsets of MoS$_2$(1L)/hBN(2L)/Gr}
\begin{figure*}[h]
\includegraphics[width=\textwidth]{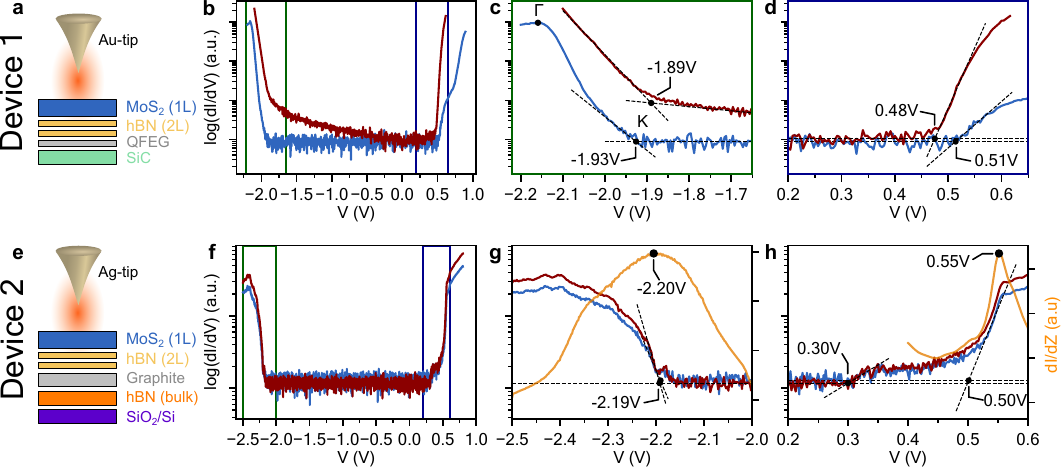}
\caption{\label{fig:SI_STS_MoS2_2L_hBN}
\textbf{Band onset estimates of MoS$_2$(1L)/hBN(2L)/Gr for Device 1 and 2.}
\textbf{a}, Schematic sketch of the heterostructure (Device 1) measured with an Au-coated tungsten tip.
\textbf{b}, STS spectra with different tip heights recorded at the same location on MoS$_2$/hBN(2L)/QFEG (Device 1) with a Au-coated tungsten tip. At closer tip-sample distance (red) the K-band onset becomes apparent but also direct tunneling to graphene contributes.
\textbf{c,d}, Close-up (boxes in b) of the valence band (c) and conduction band (d) onsets. The band onsets are estimated based on a linear fit in the semi-logarithmic plot. We find $-1.91\pm0.03$ for the valence band and $0.49\pm0.03$ for the conduction band of MoS$_2$/hBN(2L)/QFEG, resulting in a quasi-particle band gap of $2.4\pm0.04$.
\textbf{e}, Schematic sketch of the heterostructure (Device 2) measured with an Ag-tip.
\textbf{f}, STS spectra with different tip heights recorded at the same location on MoS$_2$/hBN(2L)/Gr (Device 2).
\textbf{g,h}, Estimates of the band onsets based on d$I$/d$V$ (blue, red) and d$I$/d$z$ (yellow) at constant $I$. We find $-2.20\pm0.05$ for the valence band and $0.30\pm0.05$ for the conduction band of MoS$_2$/hBN(2L)/graphite, resulting in a quasi-particle band gap of $2.5\pm0.07$.
}
\end{figure*}

\newpage
\subsection*{Saturation current Device 1 and 2}
\begin{figure*}[h]
\includegraphics[width=\textwidth]{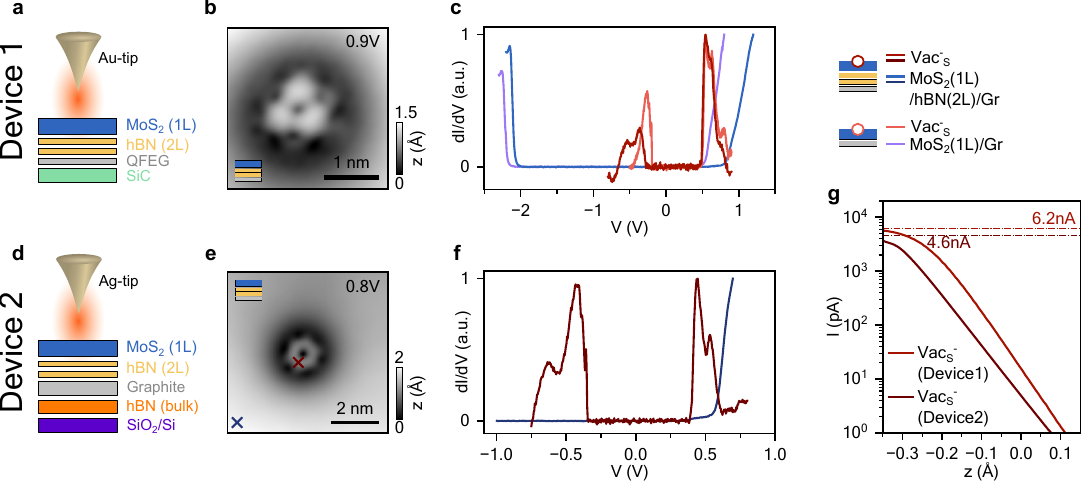}
\caption{\label{fig:SI_Sat_Current_D1_vs_D2}
\textbf{Saturation current Device 1 and 2.}
\textbf{a}, Schematic sketch of the heterostructure (Device 1) measured with an Au-coated tungsten tip.
\textbf{b}, STM topography of a Vac$_\text{S,bottom}^-$ in Device 1 (MoS$_2$/hBN(2L)/QFEG).
\textbf{c}, STS spectra of pristine MoS$_2$/QFEG (purple), pristine MoS$_2$/hBN(2L)/QFEG (Device 1, blue), and Vac$_\text{S}^-$ on these substrates.
\textbf{d}, Schematic sketch of the heterostructure (Device 2) measured with an Ag-tip.
\textbf{e}, STM topography of a Vac$_\text{S,top}^-$ in Device 2 (MoS$_2$/hBN(2L)/graphite).
\textbf{f}, STS spectra at the positions indicated in (c).
\textbf{g}, Z (height) spectroscopy ($V = -1\,$V) of Vac$_\text{S}^-$ on Device 1 and 2. In both samples the variations in saturation currents are dominated by the exact vacancy location.
}
\end{figure*}

\newpage
\subsection*{Fitting bias and current dependence of STML spectra of pristine MoS\textsubscript{2}(1L)/hBN(2L)/Gr}
\begin{figure*}[h]
\includegraphics[width=\textwidth]{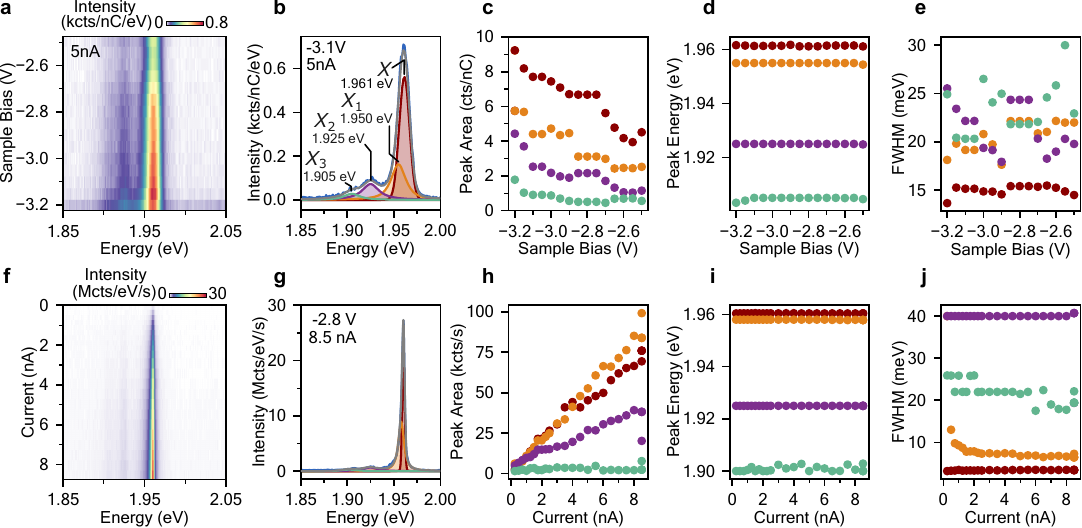}
\caption{\label{fig:SI_STML_Fitting}
\textbf{Bias and current dependence of STML spectra of pristine MoS$_2$(1L)/hBN(2L)/Gr.}
\textbf{a}, Bias dependence of STML emission on pristine MoS$_2$ using a 150\,l/mm grating at 5\,nA. 
\textbf{b}, Fit of a single spectrum at -3.1\,V by four pseudo-Voigt functions. Based on comparison to low-temperature PL data we assign the red peak to the exciton $X$ and the three peaks at lower energies to different trion species $X_{1-3}^-$.
\textbf{c}, Extracted peak areas normalized by current (cts/nC) for $X$ (red), $X_{1}^-$ (orange), $X_{2}^-$ (purple), and $X_{3}^-$ (green). All peaks increase with higher negative bias. 
\textbf{d}, Extracted peak energies for exciton and trion species. The peak energies remain essentially constant.
\textbf{e}, Extracted linewidth for exciton and trion species. The linewidths exhibit some fluctuations due to fitting instabilities. 
\textbf{f}, Current dependence of STML emission on pristine MoS$_2$ using a 600\,l/mm grating at -2.8\,V.
\textbf{g}, Fit of a single spectrum at 8.5\,nA by four pseudo-Voigt functions, corresponding to the exciton $X$ (red) and trions $X_{1-3}^-$ (orange, purple, green).
\textbf{h}, Extracted peak areas (cts/s) not normalized by current for $X$ (red), $X_{1}^-$ (orange), $X_{2}^-$ (purple), and $X_{3}^-$ (green). All peaks apepar to linearly increase as a function of injection current.
\textbf{i}, Extracted peak energies for exciton and trion species. The peak energies remain essentially constant.
\textbf{j}, Extracted linewidth for exciton (3.4\,meV) and trion species. 
}
\end{figure*}

\newpage
\subsection*{Low bias STML of pristine MoS$_{\text{2}}$(1L)/hBN(2L)/QFEG}
\begin{figure*} [h]
\includegraphics[width=0.5\textwidth]{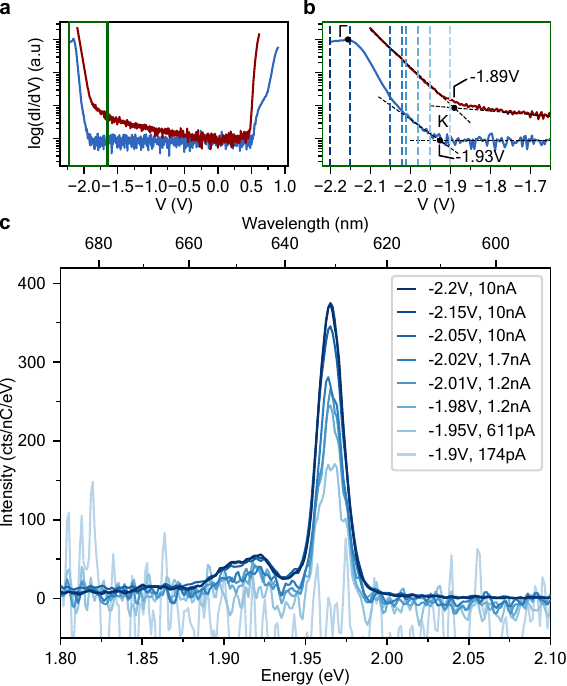}
\caption{\label{fig:SI_Low_Bias_STML_MoS2}
\textbf{Low bias STML of pristine MoS$_{\text{2}}$(1L)/hBN(2L)/QFEG}
\textbf{a}, STS spectra acquired at identical location on pristine MoS$_{\text{2}}$(1L)/hBN(2L)/QFEG using distinct tunneling setpoints (blue: far, red: close) to better resolve the onset of the K VBM with high in-plane momentum.
\textbf{b}, Magnified view of the valence band region from (a), with estimated onset energies for the K and $\Gamma$ points indicated. Vertical dashed lines denote the sample biases employed for STML measurements in (c).
\textbf{c}, STML spectra normalized by injection current recorded at the same lateral position as in (a,b), acquired under varying sample biases approaching the K-point onset. For $V < -2.05$\,V, measurements were performed in constant-height with manually approaching the tip by in order to increase tunneling current. The onset of STML emission at -1.95\,eV is concurrent with the onset of the VBM extracted from STS.
}
\end{figure*}

\newpage
\subsection*{Plasmon-absorption and plasmonic cavity}
\begin{figure*}[h]
\includegraphics[width=0.7\textwidth]{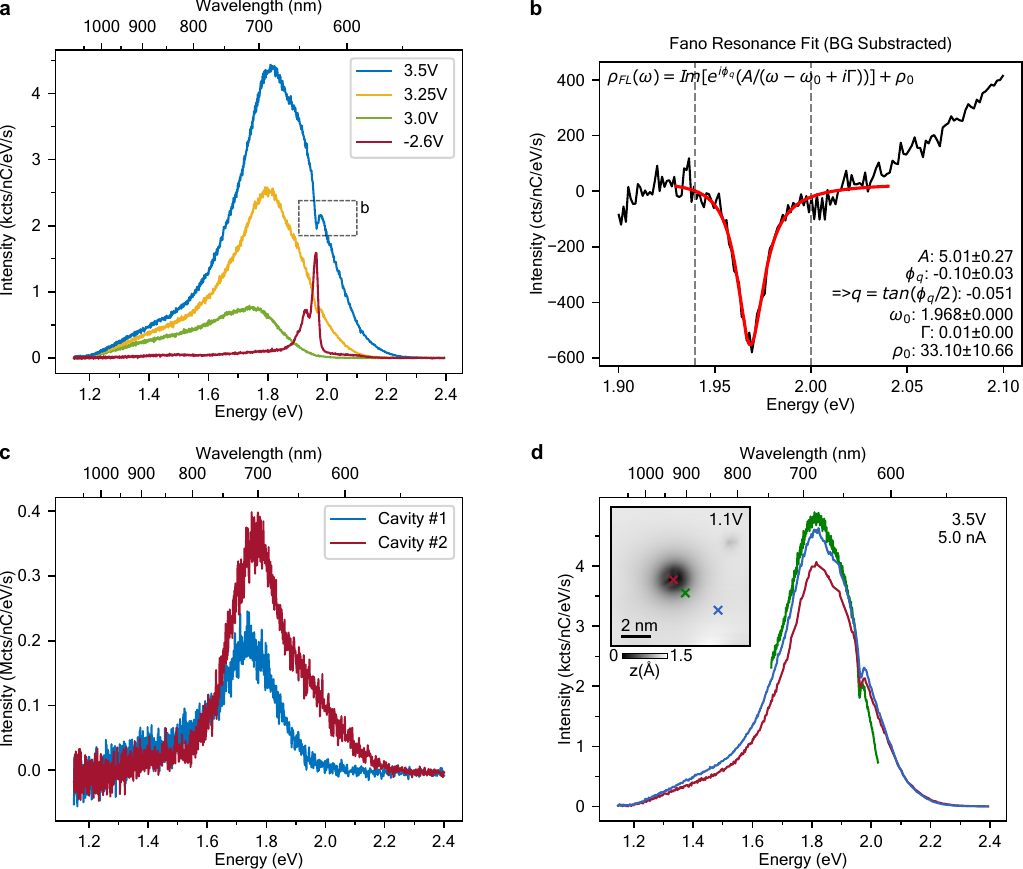}
\caption{\label{fig:SI_Plasmon}
\textbf{Plasmon absorption and plasmonic tip-sample cavity.}
\textbf{a}, STML spectra of pristine MoS\textsubscript{2}/hBN(2L)/Gr at positive and negative bias voltages. Under negative bias voltages, the pristine exciton is observed, while positive bias voltages excite the plasmon via inelastic tunneling processes. When the plasmon emission spectrally overlaps with the exciton emission, absorption leads to a dip in the plasmon spectra.
\textbf{b}, Fitting a Fano resonance to the absorption dip in the STML spectra at 3.5\,V (blue spectrum in a). Prior to fitting, a linear background was subtracted to isolate the feature. The occurrence of a Fano resonance indicates the weak coupling between the plasmonic field of the nanocavity and the emitter. 
\textbf{c}, Characterization of the tip-sample nanocavity plasmon on Au(111) at $V=3$\,V, $I=100$\,pA, 10\,s integration time, for different tips used in the experiment.
\textbf{d}, STML spectra at positive bias close to a Vac$_\text{S}^-$ defect. Apart from the plasmon absorption dip of the pristine emission, no additional dips are observed that would indicate defect emission.
}
\end{figure*}

\newpage
\subsection*{STML uniformity and comparison to photoluminescence}
\begin{figure*}[h]
\includegraphics[width=\textwidth]{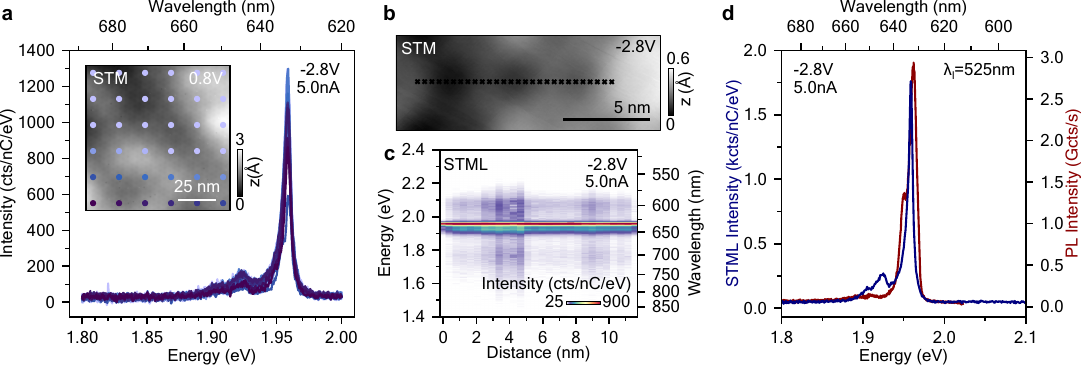}
\caption{\label{fig:SI_STML_Uniformity_and_PL}
\textbf{Pristine MoS$_2$/hBN(2L)/Gr STML emission uniformity and comparison with PL.}
\textbf{a}, Location-dependent STML spectra of pristine MoS$_2$/hBN(2L)/Gr (600\,l/mm grating). Inset shows the STM topography of a pristine area (only O$_\text{S}$ defects) with indicated STML locations.
\textbf{b}, STM topography of MoS\textsubscript{2}/hBN(2L)/Gr moiré pattern with indicated STML measurement locations from (c).
\textbf{c}, STML across the moiré in (b) as the tip passes over a depression of the moiré pattern, the broader plasmonic background intensifies.
\textbf{d}, Comparison STML and \textit{in-situ} PL emission (tip retracted) on the same sample.
}
\end{figure*}

\newpage
\subsection*{Identification of CH$_\text{S}^-$ and Vac$_\text{S}^-$}
\begin{figure*}[h]
\includegraphics[width=\textwidth]{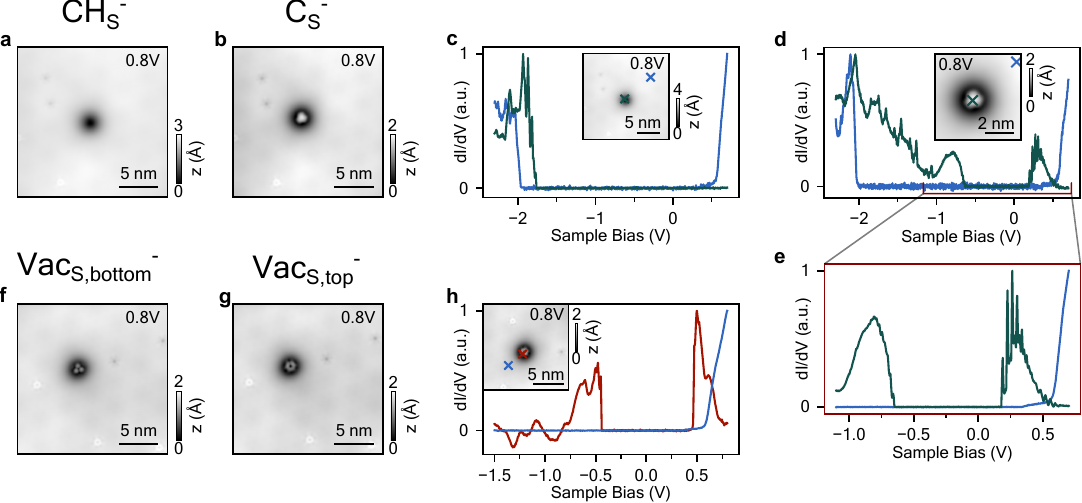}
\caption{\label{fig:SI_Identification_CHS_VacS}
\textbf{Identification of CH$_\text{S}^-$ and Vac$_\text{S}^-$.}
\textbf{a}, STM topography of the negatively charged CH$_\text{S}^-$ defect in MoS$_2$/hBN(2L)/Gr shown in Fig. 4 of the main text. 
\textbf{b}, STM topography of the same region after H dissociation with a voltage pulse (4.5\,V, 8\,nA). 
\textbf{c-e}, d$I$/d$V$ spectroscopy of CH$_\text{S}^-$ (green,c), C$_\text{S}^-$ (green,d), and pristine MoS$_2$/hBN(2L)/Gr (blue) before (c) and after (d) H dissociation at the locations indicated in the insets.   
The characteristic vibronic peaks (e) identifies the defect as negatively charged carbon impurity C$_\text{S}^-$~\cite{cochraneSpindependentvibronicresponse2021c}. 
\textbf{f,g}, STM topography of a Vac$_\text{S,bottom}^-$ (f) and Vac$_\text{S,top}^-$ (g) in MoS$_2$/hBN(2L)/Gr. Here, the vacancy switched from bottom to top, which rarely happens at high bias and high currents.
\textbf{h}, d$I$/d$V$ spectroscopy of the Vac$_\text{S,top}^-$ (red) and pristine MoS$_2$/hBN(2L)/Gr (blue) at the locations indicated in the insets.
}
\end{figure*}

\newpage
\subsection*{Exciton emission and band bending at CH$_\text{S}^-$}
\begin{figure*}[h]
\includegraphics[width=\textwidth]{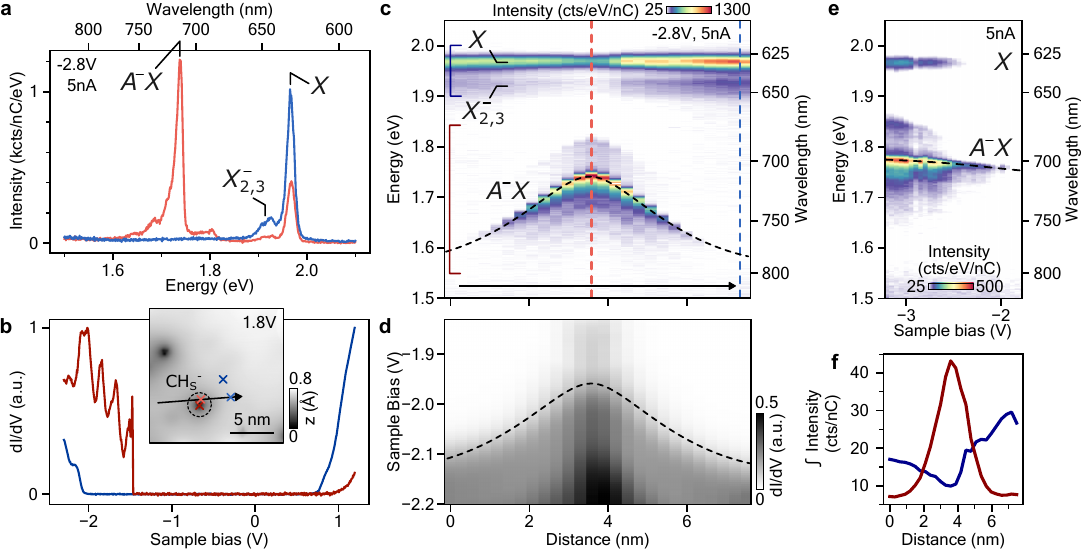}
\caption{\label{fig:SI_CD}
\textbf{Exciton emission and band bending at CH$_\text{S}^-$.}
\textbf{a}, STML spectra of pristine MoS$_2$ (blue) and negatively charged CH$^-_\text{S}$ defect (orange) recorded at -2.8\,V and 5\,nA on Device 1 (while Fig. 4 was acquired on Device 2). 
\textbf{b}, STS spectra of CH$^-_\text{S}$ (red) and pristine MoS$_{2}$ (blue). The inset shows the locations of the STML and STS measurements.
\textbf{c}, STML line spectra (off-center) across CH$^-_\text{S}$ (-2.8\,V and 5\,nA). The dashed black line follows the maximum of the defect exciton $A^-X$ emission. The locations of the STML spectra shown in (a) are indicated as dashed vertical lines in (c).
\textbf{d}, STS spectra of CH$^-_\text{S}$ recorded along the same line as the STML data in (c). The dashed black line superimposed on the valence band bending is the same as in (c), illustrating that the energy shift of the $A^-X$ emission and electronic band bending are equivalent.
\textbf{e}, STML spectra on the CH$^-_\text{S}$ defect as a function of sample bias. The onset of $A^-X$ defect emission appears at lower bias than the onset of the pristine exciton $X$ emission, indicating that direct excitation without the pristine exciton.
\textbf{f}, Integrated photon counts of STML along the line shown in (c) over the spectral range indicated by the blue and red brackets on the left. Blue: Pristine exciton and trion emission, Red: Defect emission bands.
}
\end{figure*}

\newpage
\subsection*{Current dependent STML emission of CH$_\text{S}^-$}
\begin{figure*}[h]
\includegraphics[width=\textwidth]{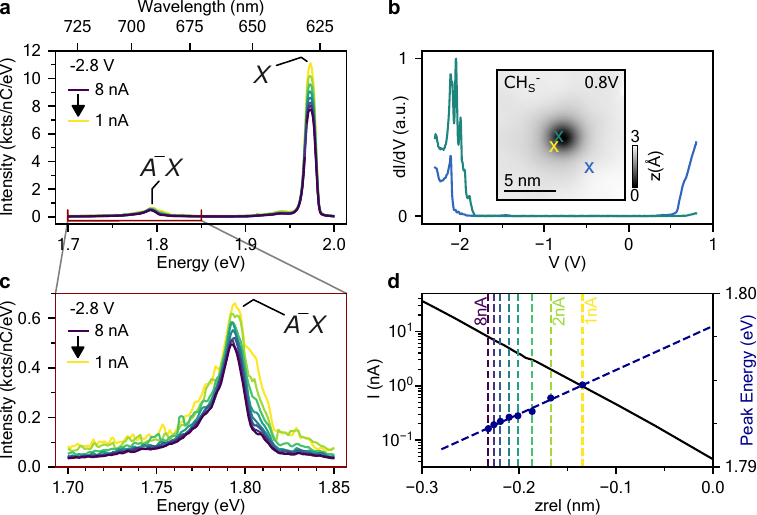}
\caption{\label{fig:SI_CurrentDep_CHS}
\textbf{Current dependent STML emission of CH$_\text{S}^-$.}
\textbf{a}, Normalized current-dependent STML emission next to CH$_\text{S}^-$ at -2.8\,V (grating 150\,l/mm). 
\textbf{b}, Constant height STS spectra of CH$_\text{S}^-$ defect (green) and pristine MoS$_2$/hBN(2L)/Gr (blue). Inset: STM topography with STS and STML (yellow marker) locations indicated.
\textbf{c}, Magnified view of normalized STML emission shown in (a) around the CH$_\text{S}^-$ $A^-X$ emission peak.
\textbf{d}, Current (black) and $A^-X$ emission peak energy (blue dots) as a function of tip height extracted from the data shown in (c). The slope of the linear fit 25\,meV/nm signifies the tip-induced Stark shift. While the tip-induced Stark shift of a few meV is noticeable, it cannot explain the giant spectral shift of the defect emission line around the charged defect.
}
\end{figure*}

\bibliography{references}